\documentclass{article}
\usepackage{epsfig,graphicx}
\usepackage{float}
\usepackage{amsmath,amsfonts}
\usepackage{authblk}
\usepackage[utf8]{inputenc}

\newcommand{\be}{\begin{equation}}
\newcommand{\ee}{\end{equation}}

\newtheorem{theorem}{Theorem}[section]
\newtheorem{definition}[theorem]{Definition}
\newtheorem{proposition}[theorem]{Proposition}
\newtheorem{lemma}[theorem]{Lemma}

\title{Analysis of trophic networks: an optimisation approach}

\author[1]{Jean-Guy Caputo\thanks{caputo@insa-rouen.fr}}
\author[2]{Valerie Girardin \thanks{valerie.girardin@unicaen.fr}}
\author[1]{Arnaud Knippel \thanks{arnaud.knippel@insa-rouen.fr}}
\author[1]{Hieu Nguyen \thanks{mingxiao13492@gmail.com}}
\author[3]{Nathalie Niquil, \thanks{nathalie.niquil@unicaen.fr}}
\author[3]{Quentin Nogu\`es, \thanks{quentin.nogues@unicaen.fr}}

\affil[1]{Laboratoire de Math\'ematiques, INSA de Rouen Normandie\\ 76801 Saint-Etienne du Rouvray, France.}
\affil[2]{UNICAEN, CNRS, Laboratoire de Math\'ematiques Nicolas Oresme, 14000 Caen, France}
\affil[3]{UNICAEN, Laboratoire Biologie des ORganismes et Ecosyst\`emes Aquatiques, FRE 2030 BOREA (MNHN, UPMC, UCBN, CNRS, IRD-207) CS 14032, 14000 Caen, France}

\date{ \today}

\begin{document}
\maketitle

\begin{abstract}
We introduce a methodology to study the possible
matter flows of an ecosystem defined by observational biomass 
data and realistic biological constraints. The flows belong to 
a polyhedron in a multi dimensional space making 
statistical exploration difficult in practice;
instead, we propose to solve a convex optimization problem. Five criteria 
corresponding to ecological network indices have been selected 
to be used as convex goal functions. Numerical results
show that the method is fast and can be used
for large systems. Minimum flow solutions are analyzed using flow
decomposition in paths and circuits. Their consistency is also tested
by introducing a system of differential equations for the biomasses and examining 
the stability of the biomass fixed point. 
The method is illustrated and explained throughout the text on an ecosystem toy model. It is also applied to realistic food models. \\

{\bf keywords : Convex optimization \and Ecosystem \and  Trophic network} \\
{\bf AMS classification:  92C42, 49N30}
\end{abstract}


\maketitle

\noindent{\bf Acknowledgements}  
This work is part of the ECUME project, co-financed by the European Union with the European regional development fund (ERDF) and by the Normandie Regional Council. 

\section{Introduction}
 
Functional ecology is based on seminal works from the XXth century
centered on the object ecosystem. From the first definition of an ecosystem by
  \cite{tansley35} to the construction of its main concepts by 
  \cite{lindeman42},
 \cite{Odum53} or  \cite{macarthur55}, among others, ecosystems have been
described as entities gathering living organisms and their habitat, and described
as dynamic entities, based on exchanges of organic matter. From those works
were derived a system analysis of these exchanges based on emergent
properties; see \cite{Odum69}, \cite{Patten76}, \cite{ula86}, \cite{frontier92}.

The description of ecosystems is often based on networks of 
interactions, of different types. For terrestrial ecosystems, 
recent developments concern different types of interactions, 
sometimes gathered into a 
common model called multiplex 
\cite{fontaine11}. In marine ecology, the most studied interactions are 
trophic, i.e. the interactions between predators and preys;
they form a network called a {\it food web}. Food webs 
in marine ecosystem are highly complex, compared to the terrestrial 
ones \cite{belgrano05}
and have been described by numerous 
models. These models have been widely used to describe the 
impact of human activities on marine ecosystems 
\cite{jorgensen11}.  They are also important tools for the sustainable 
management of marine and coastal environments \cite{langlet18}.

The trophic modeling of food webs
has been mainly based on weighted networks; see for example
the Ecopath-Ecosim-Ecospace models \cite{christensen04}.
There, each link corresponds 
to a transfer of organic matter between two trophic compartments, 
collecting individuals of similar feeding behaviors and metabolisms, 
and with the same predators. Some fluxes can be estimated using 
laboratory experiments that are often associated to field
studies, however many of them remain unknown. 
To take into account these unknown flux values
within food webs, a class of models was developed called Linear 
Inverse Modeling (LIM) \cite{niquil11}.
LIM assumes a steady state for the biomass of all compartments -- a mass balanced system.
This yields a set of linear equations (equalities)
describing the steady state or mass balance. Then,
constraints are added from field measurements of mass transfers like 
local estimations of primary production, respiration or diet contents. 
Additional constraints come from experiments or the study of other 
ecosystems. All these constraints 
constitute a set of linear equalities and inequalities 
defining a bounded multidimensional polyhedron, called a polytope, 
within which lie all realistic solutions to the problem. 
Such solutions are termed flows in graph theory.

In the literature on ecosystems, the polytope is explored using 
a random walk method, called Monte Carlo Markov Chain -- see
\cite{kones06},\cite{kones09}, \cite{vdmeersche09} and \cite{voe10} -- or
Monte Carlo Linear Inverse Modeling (MC-LIM) -- see  
\cite{kones06} and \cite{vdmeersche09}. 
Linear Inverse Monte Carlo Markov
Chain (LIM-MCMC) models are mass balanced. This stochastic approach is an indirect
way to consider the variability of the living; see \cite{niquil11}
\cite{voe10}. As such, they provide a wide range of possible
results, and not a single value like other approaches.
However, for large systems, this exploration of the polytope can be very 
long, with a very large number of simulation runs.

Gathering indices from various domains, from information theory to input- output analysis in econometry, several Ecological Network Analysis  indices have been introduced for describing the organization of the flows and the functioning of the ecosystem \cite{latham06}, or as criterions of ecosystem maturity \cite{ula97}; One such simple index is the sum of the flow components squared
\cite{vezina88}; see \cite{jdd14}, \cite{kones09}, \cite{stbeat13},
and \cite{ula18} for many others.
These indices assume that the flows are given, but LIM results can be used to compute approximated values.

In this article, we propose to use these indices as 
goal functions and combine them with the constraints 
to set up an optimization problem. This procedure 
has a low computing time compared to the LIM MC-MC method and 
directly yields a unique flow solution within the polytope, if 
the goal function is convex. Note that \cite{stbeat13} combined the LIM-MCMC exploration of the polytope to comparison of  different indices
to select a unique flow vector.

\begin{figure} 
\begin{center} 
\epsfig{file=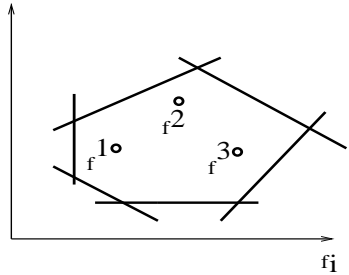,height=4cm,width=5.5cm}
\end{center}
\caption{Schematic drawing of the polytope with the
optima for  different goal functions.
} 
\label{schema}
\end{figure}

Fig. \ref{schema} presents a 
schematic picture of the polytope in flow space $f$ together
with three minima corresponding to three convex goal functions.
To compare the optima obtained through optimization of goal functions, say $f^1,~f^2$  and $f^3$, a first step is to examine
the main flows from an ecological point of view and see if they appear
reasonable. In a more quantitative approach, one can decompose these
flows into paths and circuits and again check these using ecology
know-how. We also suggest, using simple rules, to introduce a 
dynamical system satisfied
by the biomasses and whose coefficients depend on the flow solution
$f^1,~f^2$  or $f^3$. This dynamical system has a fixed point -- the 
given biomasses, whose  stability can be determined. If the fixed point is stable, then the model is consistent, say for example
$f^2$ in Fig. \ref{schema}. Then the optimum $f^2$ yields
an acceptable solution to describe the ecosystem.

Using a six species toy model inspired by a realistic ecosystem,
we proceed to illustrate this methodology. We do not pretend 
that the model is realistic but we focus on the analysis and present it
in as much detail as possible. This detailed presentation is easy
to follow on the six species system and naturally extends 
to an ecosystem of any size. 
We write explicitly the constraints defining the polytope in Section
2. In Section 3, we discuss flow decomposition into paths and
circuits, a general result from the theory of polyhedra.
The full optimization problem is presented in Section 4.
First we present the five convex goal functions, three of which 
are independent of the constraints and 
two depend on the constraints. 
The results of the optimization problem are 
analyzed using the flow decomposition of Section 3. 
From these flow solutions, we
write the formal dynamical system for the biomasses and examine
the stability of the fixed point in Section 5. We show that the fixed point
is always marginally stable, in the absence of detritus, in Section 6.
We show that the detritus controls the stability and give a sufficient
condition for the fixed point to be stable. Conclusions and application to real data are presented in
Section~7.

\section{The model and notation}\label{model6species}

A realistic model of a marine ecosystem with nineteen species 
was introduced and analyzed by the authors in \cite{nogues20}.
To focus on the method of analysis, we simplified this model
and reduced it to an ecosystem of six species. This methodology 
can be extended to ecosystems or arbitrary size;  
in Section~6, we give some results for the realistic ecosystem 
studied in \cite{nogues20}.

\begin{figure}
\begin{center}
  \includegraphics[scale = 0.6]{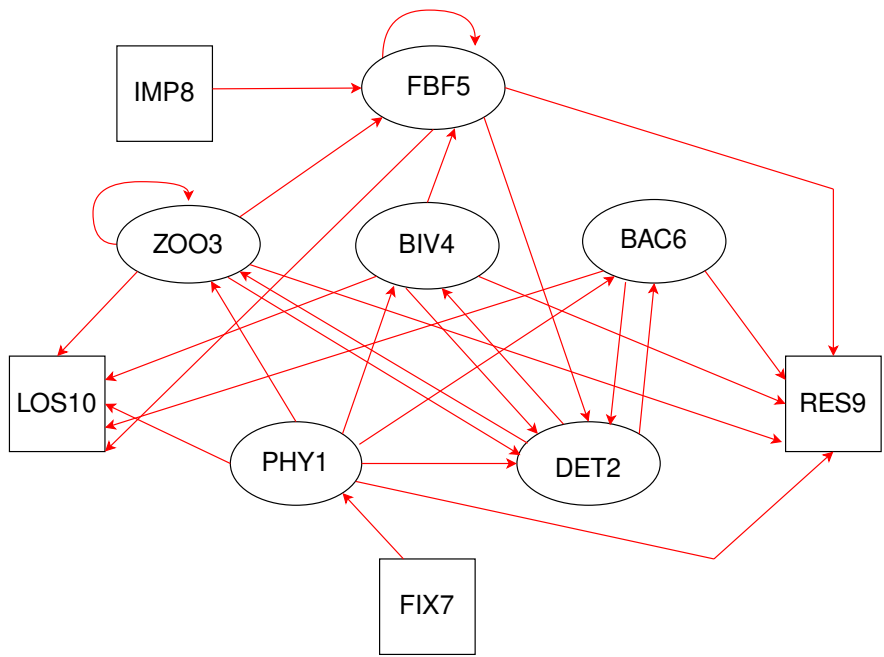}
\caption{The 6-species model ecosystem described in Section \ref{model6species}.}
\end{center}
\label{eco6}
\end{figure}

The graph of this simplified 6-species ecosystem   is presented in Fig.~\ref{eco6}. 
The ordered types of living organisms with biomasses --   circles  in Fig.~\ref{eco6}  -- are:
$$
  \begin{array}{l}
    \mbox{Phytoplankton} \equiv \mbox{PHY1},
\quad  \mbox{Zooplankton} \equiv \mbox{ZOO3 },\quad
    \mbox{ Bivalve} \equiv\mbox{BIV4}, \\
     \mbox{Fish  benthic  feeders} \equiv\mbox{FBF5},
   \quad \mbox{Bacteria} \equiv\mbox{BAC6}.
\end{array}
$$
The other vertices --  rectangular boxes in Fig.~\ref{eco6}  -- are: 
$$
  \begin{array}{l}
     \mbox{Detritus} \equiv\mbox{DET2} ,
    \quad \mbox{Photosynthesis  of  phytoplankton} \equiv\mbox{FIX7}, \\
     \mbox{Respiration} \equiv\mbox{RES9}, \quad
    \mbox{Fishing, Trawl, Dredge,...} \equiv\mbox{LOS10},\\
       \mbox{Import to system (ability of a species to move geographically}\\ \hphantom{Import to system (ability }\mbox{in order to feed)} \equiv\mbox{IMP8} .
\end{array}
$$
The arrows between the compartments (edges between vertices) represent matter 
flows that are inferred by ecologists. Following graph theory, terminology
such oriented edges will be denoted arcs.

 We denote by
$S$ the set of all vertices  including the detritus   DET2, 
$$ S=\{PHY1, DET2, ZOO3, BIV4, FBF5, BAC6\},
$$ 
 with
$S'=S -\{DET2\}$ the set of all species vertices, and
$E$ the set of all arcs -- denoted by $ij$ when going from 
vertice $i$ to vertice $j$.

The central mathematical object of this article is the flow.
\begin{definition} Let the successors and predecessors of a vertice $i$ be
$N^+(i) =  \{\mbox{vertices} ~j,~~ ij \in E \}$ and $
N^-(i) =  \{\mbox{vertices} ~j,~~ ji \in E \}.$

A flow is a vector, of dimension the number of arcs with non negative components,
satisfying Kirchoff law on all vertices of $S$,
\begin{equation}\label{kirchoff}
     \sum_{j\in N^+(i) } f_{i,j}  -  \sum_{j\in N^-(i)} f_{j,i} = 0,\quad   i\in S,
\end{equation}

\end{definition}

In many situations, all biomasses $B_i$ of the species $i$ can be 
measured with accuracy. On the other hand, the flow components 
$f_{i,j}$  are much more difficult to evaluate. 
Therefore, we will adopt here the standard viewpoint 
that the biomasses are given and the flows between nodes are unknown.

\subsection{Biological constraints}

The flow components $f_{i,j}$ between the compartments satisfy 
biological constraints. For example, a fish cannot eat 
more that a certain percentage of its biomass. 
When defining the constraints, we gather all available information, if 
possible from studies of the local ecosystem, if not, from 
ecosystems similar to ours. In absence of information, 
the constraints are derived from experiments or from empirical 
equations.

\begin{definition}
For all species $i\in S$, the production $P_{i}$   is  
\begin{eqnarray*} \label{prod}
& P_{i} = \sum_{j\in S'} f_{j,i}  - f_{i,{\rm res}} - f_{i,{\rm det}}, 
\end{eqnarray*}
where $f_{i,\rm{res}}$ is the respiration flow   and $f_{i,{\rm det}}$ is
the excretion flow   -- assuming it goes to  the detritus.
\end{definition}

Among non species compartments (vertices), the detritus plays a singular
role as the only one for which 
flows go in and out of. It is then natural to assume
for it a Kirchoff law where in-going equal out-going flows.
For the species vertices in $S'$,   the following constraints are imposed by
biological observations through nonnegative coefficients $c$. 
These coefficients come from field measurements
of mass transfers like local estimations of primary production, respiration or
diet contents. They can also be estimated from experiments or the study of
other ecosystems.

\begin{description}
\item{\bf Positivity of the flow components} \ $ f_{i,j} \geq 0 ,$ for all $ ij \in E.
$

\item{\bf Kirchoff law at the vertices} 
Equations \eqref{kirchoff} express the conservation of mass at each species vertex.

\item{\bf Primary production constraint}  
The production $P_1$ of the entry in the ecosystem 
Phy1 is bounded,
$$ c^{-}_{\rm pro}  \leq P_{1} \leq c^{+}_{\rm pro}. $$
This constraint comes from a local study estimating the carbon incorporated, with enrichments in 13Cu a stable isotope of carbon, compared with studies based on an estimation of the activity of photosystems using pulse amplitude modulation \cite{nc12}.

\item{\bf Respiratory constraints} 
The respiration flow $f_{i,{\rm res}}$ of each species $i$ is bounded,
$$ c^{-}_{{\rm res},i}\sum_{j\in N^-(i)} f_{j,i}  \leq f_{i,{\rm res}} \leq c^{+}_{{\rm res},i}\sum_{j\in N^-(i)} f_{j,i}, \quad \mbox{for all }  i\in S'. $$

\item{\bf Excretion constraints} 
The excretion of all species $i$  but the phytoplankton is bounded,
$$c^{-}_{{\rm det},i}\sum_{j\in N^-(i)} f_{j,i}  \leq f_{i,{\rm det}} \leq c^{+}_{{\rm det},i}\sum_{j\in N^-(i)} f_{j,i}, \quad \mbox{for all }  i\in S', i \neq {\rm PHY1}.
$$
The phytoplankton excretion is bounded too,
$$  c^{-}_{{\rm det},{\rm phy}} P_{{\rm phy}}  \leq f_{{\rm phy},{\rm det}} \leq c^{+}_{{\rm det},{\rm phy}} P_{{\rm phy}}.
$$

\item{\bf Food conversion efficiencies constraints} 
The production of a species $i$ is constrained by bounds  depending on the entering flow of species $i$
$$
    c^{-}_{{\rm eff},i}\sum_{j\in N^-(i)} f_{j,i}  \leq P_{i} \leq c^{+}_{{\rm eff},i}\sum_{j\in N^-(i)} f_{j,i}, \quad \mbox{for all }  i\in S'.
$$

\item {\bf Production related to biomass constraints} 
The production of a species $i$ is constrained by bounds  depending on   its biomass $B_{i}$,
$$
    c^{-}_{{\rm bio},i}B_{i}  \leq P_{i} \leq c^{+}_{{\rm bio},i}B_{i}, \quad \mbox{for all }  i\in S',
$$

\item {\bf Diet constraints} 
The entry flow $f_{j,i}$ for species $i$  is constrained by bounds  depending on   the sum of the entry flow of this species,
$$ c^{-}_{{\rm diet},i}\sum_{j\in N^-(i)} f_{j,i}  \leq f_{j,i} \leq c^{+}_{{\rm diet},i}\sum_{j\in N^-(i)} f_{j,i}, \quad \mbox{for all }  i\in S'.
$$
Diet information is derived from stomach content analyses; see e.g.  \cite{ddd05}.

\end{description}

Different empirical equations 
are gathered to define relationships between production and biomass 
or consumption and biomass, or respiration and consumption; see  \cite{brey01}, \cite{Froese04}. These equations use the shape of the caudal  fin, the individual weight, temperature, growth, etc. The individual mass 
and total biomass per $\mbox{km}^2$ values are estimated from local 
field studies \cite{dr08} or field studies from a 
similar ecosystem if not available; see, e.g., for zooplankton \cite{re03}.

All above constraints can be summarized in the following set of equations, where  $f_i$ is the total flow entering a species $i$.
\begin{eqnarray} \label{contr}
    & f_{i,j} \geq 0, & ij \in E, \\
    & f_i = \sum_{j\in N^-(i)} f_{j,i}, &   i\in S, \\
    & \sum_{j\in N^+(i)} f_{i,j}  -  \sum_{j\in N^-(i)} f_{j,i} = 0, &  i \in S, \\
    & f_1 - f_{1,{\rm res}} - f_{1,{\rm det}} - c^{+}_{{\rm pro}} \leq 0, \\
    & - f_1 + f_{1,{\rm res}} + f_{1,{\rm det}} +  c^{-}_{{\rm pro}} \leq 0 , \\
    & -c^{+}_{{\rm res},i} f_i + f_{i,{\rm res}}   \leq 0 , &   i\in S',  \\
    & c^{-}_{{\rm res},i} f_i - f_{i,{\rm res}}   \leq 0 , &    i\in S',\\
    & -c^{+}_{{\rm det},i}f_{i} + f_{i,{\rm det}} \leq 0 , &    i\in S', i \neq \mbox{PHY1}  \\
     & c^{-}_{{\rm det},i}f_{i}  - f_{i,{\rm det}} \leq 0 , &   i\in S', i \neq \mbox{PHY1} \\
    & -c^{+}_{{\rm det},{\rm phy}}P_{{\rm phy}} + f_{{\rm phy},{\rm det}} \leq 0 ,\\
    & - f_i + f_{i,{\rm res}} + f_{i,{\rm det}} +  c^{-}_{{\rm eff},i}f_{i} \leq 0 , &   i\in S',      
   \\
    &f_i - f_{i,{\rm res}} - f_{i,{\rm det}}  - c^{+}_{{\rm bio},i}B_{i}  \leq 0 , &   i\in S',\\
    &-f_i + f_{i,{\rm res}} + f_{i,{\rm det}}  + c^{-}_{{\rm bio},i}B_{i}  \leq 0 , &   i\in S',\\
    & f_{j,i} -c^{+}_{{\rm diet},i,j} f_{i}  \leq 0 , & j\in S,   i\in S', \\
    & - f_{j,i} + c^{-}_{{\rm diet},i,j} f_{i} \leq 0, & j\in S,    i\in S'. \label{contrfin}
\end{eqnarray}

For the 6-species ecosystem defined in Fig. \ref{eco6}, these 
constraints lead to the bounds on the flows shown in Table \ref{tab1}.  
More precisely, for example, the bounds on $f_{1,2}$  are obtained by
minimizing or maximizing $f_{1,2}$ together with the constraints
\eqref{contr} to \eqref{contrfin}.
\begin{table} 
\centering
\begin{tabular}{|l|r|}
 \hline
 Components & bounds \\
 \hline
 $f_{1,2}$ & [10.300 ; 51.600]  \\
 $f_{1,3}$ & [0.000 ; 136.181]  \\
 $f_{1,4}$ & [35.818 ; 172.000]  \\
 $f_{1,6}$ & [0.000 ; 96.498]  \\
 $f_{1,9}$ & [5.963 ; 95.828]  \\
 $f_{1,10}$ & [0.000 ; 129.668]  \\
 $f_{2,3}$ & [0.000 ; 27.892] \\
  $f_{2,4}$ & [9.768 ; 92.615]  \\
 $f_{2,6}$ & [0.000 ; 271.796]  
\\ \ & \\
 \hline
\end{tabular}
 \begin{tabular}{|l|r|}
 \hline
 Components & bounds \\
 \hline
 $f_{3,2}$ & [0.000 ; 82.036]  \\
 $f_{3,3}$ & [0.000 ; 16.407]  \\
 $f_{3,4}$ & [0.000 ; 39.002]  \\
 $f_{3,5}$ & [0.000 ; 35.445]  \\
 $f_{3,9}$ & [0.000 ; 49.222]  \\
 $f_{3,10}$ & [0.000 ; 78.113]  \\
 $f_{4,2}$ & [11.722 ; 111.138]  \\
 $f_{4,5}$ & [0.000 ; 35.445]  \\
 $f_{4,9}$ & [18.235 ; 197.447]  
 \\  $f_{4,10}$ & [0.000 ; 78.260]  \\
   \hline
\end{tabular}
 \begin{tabular}{|l|r|}
 \hline
 Components & bounds \\
 \hline

 $f_{5,2}$ & [0.638 ; 35.445]  \\
 $f_{5,5}$ & [0.000 ; 6.380]  \\
 $f_{5,9}$ & [3.190 ; 31.901]  \\
 $f_{5,10}$ & [0.000 ; 6.380]  \\
 &\ \\
 \hline
\end{tabular}
 \begin{tabular}{|l|r|}
 \hline
 Components & bounds \\
 \hline

 $f_{6,2}$ & [0.000 ; 112.581]  \\
 $f_{6,9}$ & [0.000 ; 180.966]  \\
 $f_{6,10}$ & [0.000 ; 192.996]  \\
 $f_{7,1}$ & [119.263 ; 319.428]  \\
 $f_{8,5}$ & [0.000 ; 35.445]  \\
 \hline
\end{tabular}
\caption{Bounds on the flow components 
for the 6-species ecosystem of Fig. \ref{eco6}.}
\label{tab1}
\end{table}

\section{Flow decomposition}

Since the constraints  \eqref{contr} to \eqref{contrfin} are all linear, the defined domain is polyhedral.
A polyhedron in $\mathbb{R}^n$ is an intersection of a finite number of 
half-spaces, in other words $P = \{x \in\mathbb{R}^n  |Ax \leq b \}$, where
 $A$ is an $n' \times n$ matrix with $n'> n$ and $b \in\mathbb{R}^n $. A bounded polyhedron is called a polytope.

The  decomposition $P = Q + C$ for all polyhedrons $P$, with $Q$ a polytope 
and $C$ a cone, is classical; see Nemhauser and Wolsey \cite{nw88}. 
Since $Q$ and $C$ are convex sets, any point in $P$ can be expressed as the
sum of a convex combination of the extreme points of $Q$ (the vertices) 
and a combination of the extreme rays of $C$ with non negative
coefficients :
\be\label{chemcycle} 
x= \sum_{i\in I} \alpha_i q^i + \sum_{j\in J} \beta_j r^j ,
\ee
where $I$ is the index set of the vertices of $Q$, $J$  the index
set of the extreme rays of $C$ and the coefficients $\alpha_i$ satisfy
$ \sum_{i\in I} \alpha_i =1$ and $\alpha_i, \beta_j$ are non negative.

 For webs of flows, the vertices of $Q$ are indicator vectors of elementary
paths 
and the extreme rays are -- up to a constant coefficient -- indicator 
vectors of elementary circuits of the web (paths that begin and 
ends at the same vertex). Therefore any numerical solution of such 
an ecological system can be
interpreted in terms of paths and circuits. 
A linked important notion is the flow value.  
\begin{definition}
For a given path of circuit, the flow value is the smallest flux
for all arcs of the path or circuit. 
\end{definition}
Usually, the decomposition (\ref{chemcycle}) is not unique. 
However, the largest flow values of the network will make some paths or circuits  necessary in any decomposition, and hence important for interpreting the obtained ecosystem solution.
An algorithm to find all necessary paths or circuits is the following:

\noindent Repeat 
\begin{itemize}
\item Find the path or circuit with largest value $\alpha$;
\item Take out $\alpha$ from the flow components of the arcs of this
path or circuit;
\end{itemize}
until stopping criterion is met.

Such an analysis assumes that  the numerical values of the flows are 
known and satisfy the biological constraints. Unfortunately, such 
exact numerical values are generally unavailable, and their a priori approximated  values do not satisfy the constraints especially the conservation  ones. The next section presents an 
optimization approach addressing this issue. A flow solution is 
computed from the knowledge of only the biomasses and some 
approximated values of the flows, or intervals of approximated values. Then,     examples of paths and circuits  extracted using the above method of flow decomposition are given for the 6-species network.

\section{The Optimization problem } 

\subsection{Goal functions}\label{goal}

We consider five  goal functions,  the most classical least squares,  Ecological Network Analysis indices, and  functions adapted from information theory. All are convex, yielding a unique optimum corresponding to a unique functioning state of the system. An important characteristics is whether the goal function includes information from the constraints or not. This separates the functions
in two classes.

In one class, no information is used from the constraints.
The first function corresponds to the quadratic energy, the classical 
least squares method,  and the solution will be the minimum of
\begin{equation}\label{F1}
     F_{1}(f) = \sum_{ij\in E} f_{i,j}^{2} .
\end{equation}
The second function is minus Shannon  entropy from information theory \cite{ct2006}, introduced  for ecological systems in 1955 in \cite{macarthur55}, 
\begin{equation}\label{F2}
F_{2}(f)
   = \sum_{ij\in E} p_{i,j}\ln(p_{i,j}) =
    \sum_{ij\in E} \frac{f_{i,j}}{f_{..}}\ln\left(\frac{f_{i,j}}{f_{..}}\right),
\end{equation}
where  the sum of all flows is
$ f_{..}={\sum_{ij\in E} f_{i,j}} ,
$
and the proportion of flows from vertex $i$ to vertex $j$ is
\be\label{pij} 
 p_{i,j} = \frac{f_{i,j}}{f_{..} }.
\ee
Classically, entropy is a concave function and is to be maximized. Here, for practical purposes,
we only deal with convex goal functions and therefore changed the
sign. With this sign, $F_{2}$ is convex -- see Appendix 2.

Finally, the third function is minus the system 
redundancy (overhead) introduced  in 
 \cite{ula97} as
\begin{equation}\label{F3}
F_{3}(f)=
\sum_{ij\in E} p_{i,j}
\ln\left(\frac{p_{i,j}}{p_{i.}}\frac{p_{i,j}}{p_{.j}}\right) = 
\sum_{ij\in E}  \frac{f_{i,j}}{f_{..}}\ln\left(\frac{f_{i,j}^2}{f_{i.}f_{.j}}\right),
\end{equation}
where the marginal proportions are  
$$p_{i.}={\sum_{j\in S :~ij\in E} p_{i,j}}\quad \mbox{and}\quad p_{.j}={\sum_{i\in S :~ij\in  E} p_{i,j}} ;
$$
note that the sum is on all $i$ or $j$ such that $ij$ is an arc of $E$.
As for the entropy, with this choice of sign, $F_{3}$ is convex  -- see Appendix 2.

In information theory terms, $-F_2$ is the Shannon entropy of the system and 
$-F_3$ is a symmetrized conditional entropy; see 
 \cite{ct2006}  and also 
 \cite{ula18} for details on such  entropic indexes.
Note that, although it is a well-known Ecological Network Analysis index, we do not consider the ascendency of 
\cite{hu84} 
because it is neither convex nor concave -- see Appendix 2.

In the other class, the goal functions  incorporate information from the constraints.
The most classical quantity in this aim in information theory is the Kullback-Leibler divergence 
introduced in \cite{kullback51}, that measures a "distance" between  two distributions,
\begin{equation}\label{F4}
 F_{4}(f) = K(f|f^*)=  
\sum_{ij\in E} p_{i,j}
\ln\left(\frac{p_{i,j}}{p^*_{ij}}\right) ,
\end{equation}
where the proportions  are given in \eqref{pij}.
The divergence is not a mathematical distance because it is not symmetric in $p$ and $p^*$. Still, it is nonnegative and null only if $p=p^*$, and minimizing $K$ determines the projection in terms of divergence 
of the reference   $f^*$ (or $p^*$) on the set of solutions $f$ to the constraints; see \cite{ct2006} and \cite{csiszar75}.
 A natural way, that makes sense in ecology, to include the information from the constraints is to set all $f^*_{ij}$ as  the middle of the constraint intervals
$[{f_{ij}}^{\min}, {f_{ij}}^{\max} ]$.
Note that  $F_2$ is  the Kullback-Leibler 
divergence where $f^*$ is the uniform distribution on the flows. 
This uniform distribution is usually not a flow; this
confirms the importance of using graph theory to describe such systems.

Finally, a simple generalization of the quadratic function $F_1$ is 
\begin{equation}\label{F5}
F_{5}(f) = \sum_{ij\in E} (f_{i,j}-f_{i,j}^*)^{2} .
\end{equation}

\subsection{The full optimization problem}\label{4.2sec}

Combining the biological constraints (\ref{contr}) 
to (\ref{contrfin}) with the goal
functions \eqref{F1} to \eqref{F5} yields the well posed convex
optimization problem in a positive 
polytope of $\mathbb{R}^n$,
\begin{eqnarray} \label{fullopt}
    && \min_{f} F ,\\
    && {\rm with~constraints} ~~(\ref{contr})-(\ref{contrfin}) , \nonumber
\end{eqnarray}
where $F$ can be any one of the five goal functions given in \eqref{F1} to \eqref{F5}.

A well suited method to solve (\ref{fullopt})  is the 
Sequential Quadratic Programming (SQP) which uses an Augmented Lagrangian 
Solver;  see \cite{sqp} contained in
the R library NlcOptim   \cite{NlcOptim}. 
We used the R software infrastructure for the optimization. The 
algorithm is presented in Appendix 1.

\begin{table} 
\centering
\begin{tabular}{|p{1.5cm}|p{1.3cm}|p{1.3cm}|p{1.3cm}|p{1.3cm}|p{1.3cm}|}
 \hline
 Component & Solution 1 & Solution 2 & Solution 3  & Solution 4 & Solution 5 \\
 \hline
 $f_{1,2}$ & 10.300 & 10.300 & 10.940 & 15.924 & 39.948 \\
 $f_{1,3}$ & 25.469 & 45.800 & 41.130 & 44.577 & 51.098 \\
 $f_{1,4}$ & 38.784 & 36.175 & 35.818 & 55.847 & 83.598  \\
 $f_{1,6}$ & 5.272 & 13.237 & 14.186 & 25.257 & 27.971 \\
 $f_{1,9}$ & 5.963 & 7.907 & 16.766 & 18.026 & 29.146 \\
 $f_{1,10}$ & 33.472 & 7.787 & 12.040 & 21.268 & 9.331 \\
 $f_{2,3}$ & 4.499 & 7.722 & 6.576 & 6.515 & 10.465 \\
 $f_{2,4}$ & 17.814 & 23.020 & 22.668 & 26.501 & 52.497 \\
 $f_{2,6}$ & 12.302 & 50.156 & 33.101 & 70.547 & 140.374 \\
 $f_{3,2}$ & 5.993 & 12.382 & 9.910 & 19.171 & 30.782 \\
 $f_{3,3}$ & 0.000 & 1.658 & 1.847 & 2.614 & 0.000 \\
 $f_{3,4}$ & 8.525 & 6.577 & 6.638 & 8.270 & 15.900 \\
 $f_{3,5}$ & 3.244 & 11.057 & 9.144 & 5.068 & 5.322 \\
 $f_{3,9}$ & 8.990 & 15.208 & 14.866 & 7.682 & 6.156 \\
 $f_{3,10}$ & 3.213 & 8.297 & 7.147 & 10.899 & 3.402 \\
 $f_{4,2}$ & 16.542 & 18.746 & 18.165 & 27.340 &  58.398 \\
 $f_{4,5}$ & 4.508 & 11.591 & 10.462 & 6.327 & 10.692 \\
 $f_{4,9}$ & 29.045 & 27.294 & 27.421 & 41.977 & 74.061 \\
 $f_{4,10}$ & 15.028 & 8.212 & 9.075 & 14.973 & 8.845 \\
 $f_{5,2}$ & 0.901 & 21.596 & 10.197 & 12.091 & 22.394 \\
 $f_{5,5}$ & 1.232 & 3.834 & 3.513 & 2.278 & 6.380 \\
 $f_{5,9}$ & 5.681 & 24.845 & 16.380 & 8.072 & 16.014 \\
 $f_{5,10}$ & 1.201 & 2.545 & 2.866 & 1.741 & 0.000 \\
 $f_{6,2}$ & 0.878 & 17.874 & 13.131 & 29.036 & 51.814 \\
 $f_{6,9}$ & 8.384 & 22.759 & 20.390 & 32.476 & 55.258 \\
 $f_{6,10}$ & 8.384 & 22.759 & 13.766 & 34.310 & 61.273 \\
 $f_{7,1}$ & 119.263 & 121.207 & 130.883 & 180.919 & 241.095 \\
 $f_{8,5}$ & 0.031 & 26.411 & 9.838 & 10.508 & 22.394 \\ 
 \hline
\end{tabular}
\caption{Flow solutions for the 6-species system, obtained for the   goal
functions $F_1$ to $F_5$.}
\label{tab2}
\end{table}

\begin{figure}
\begin{center}
\fbox{\epsfig{file = 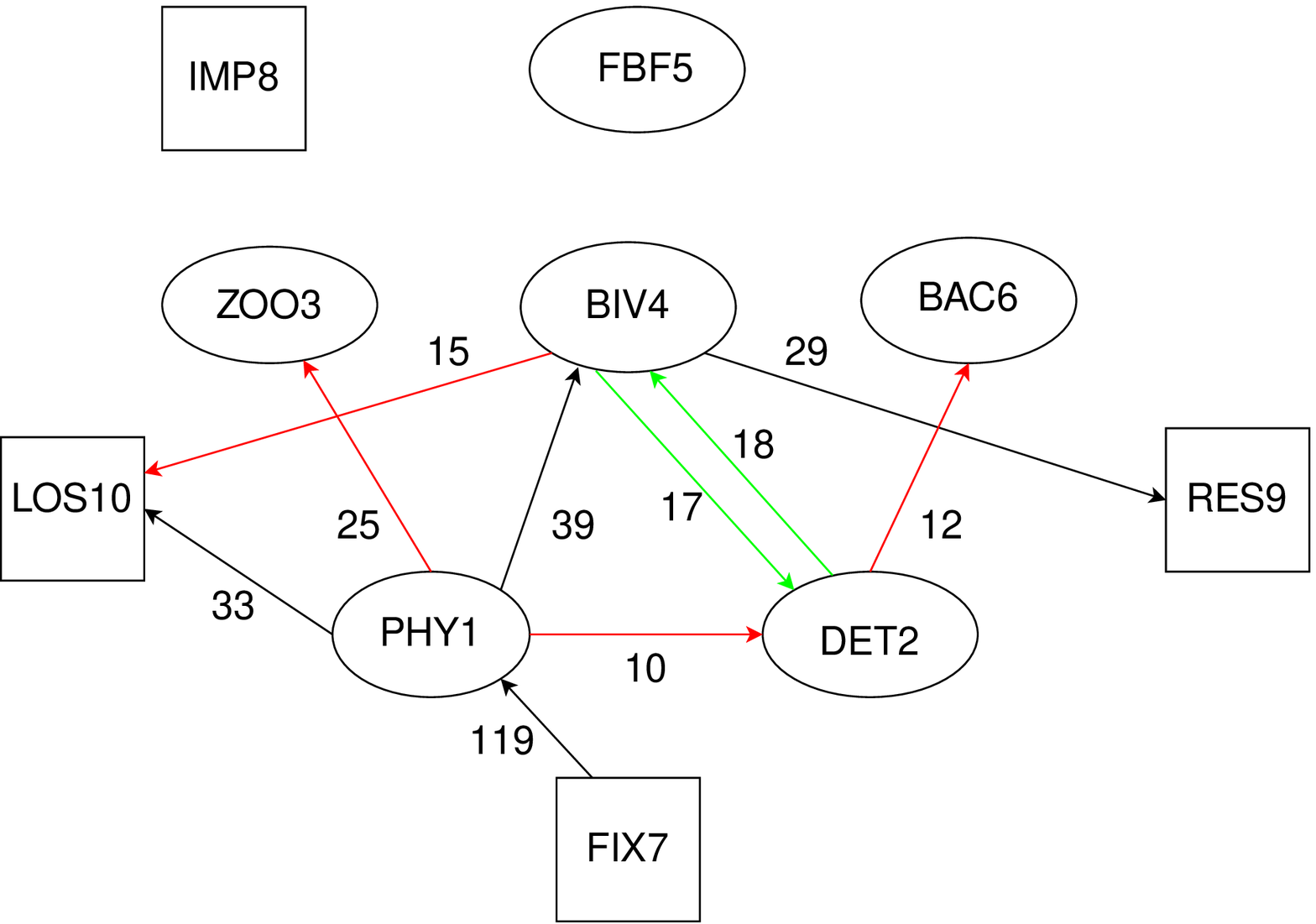,height=6cm,width=5cm}}
\fbox{\epsfig{file = 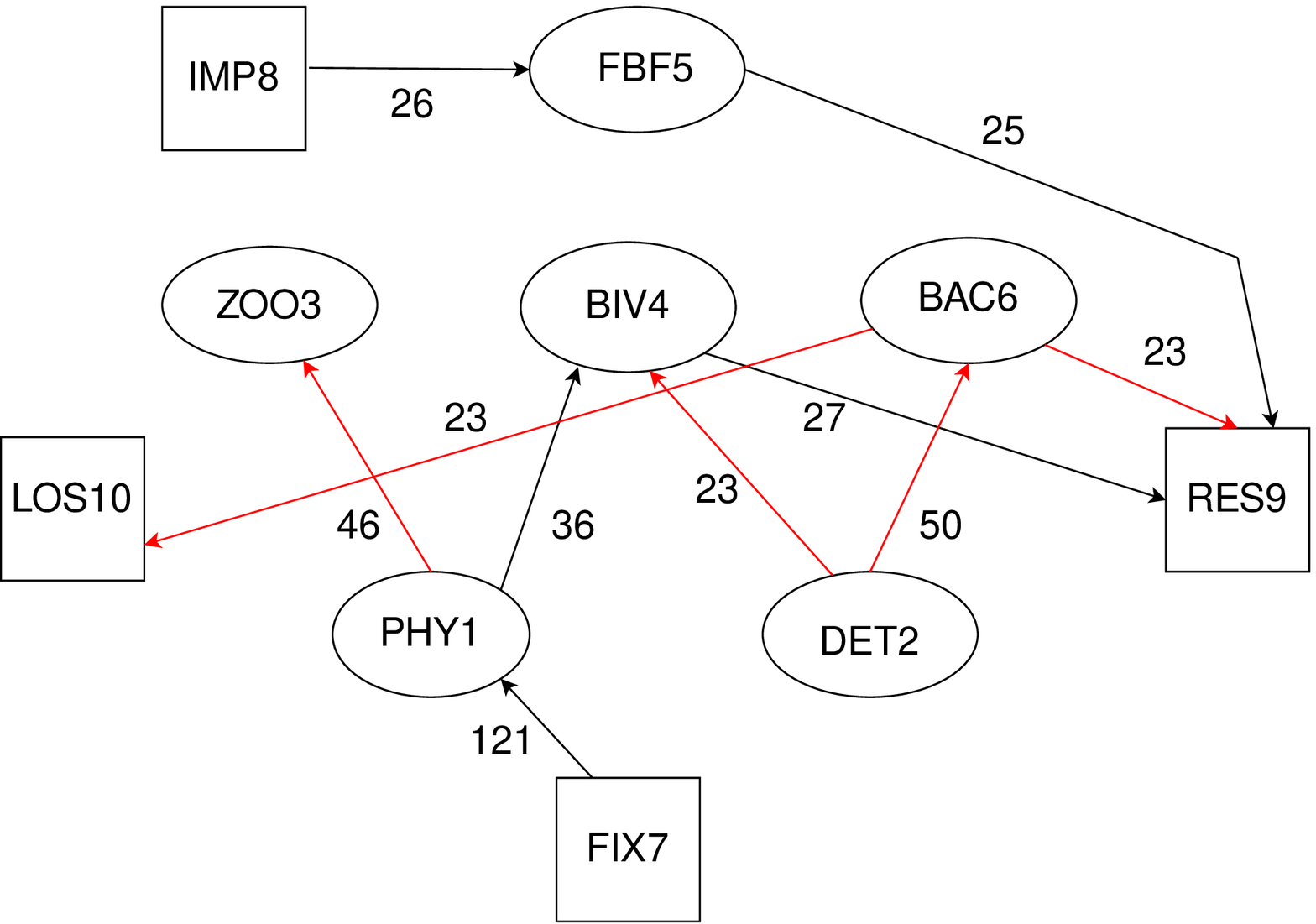,height=6cm,width=5cm}}
\fbox{\epsfig{file = 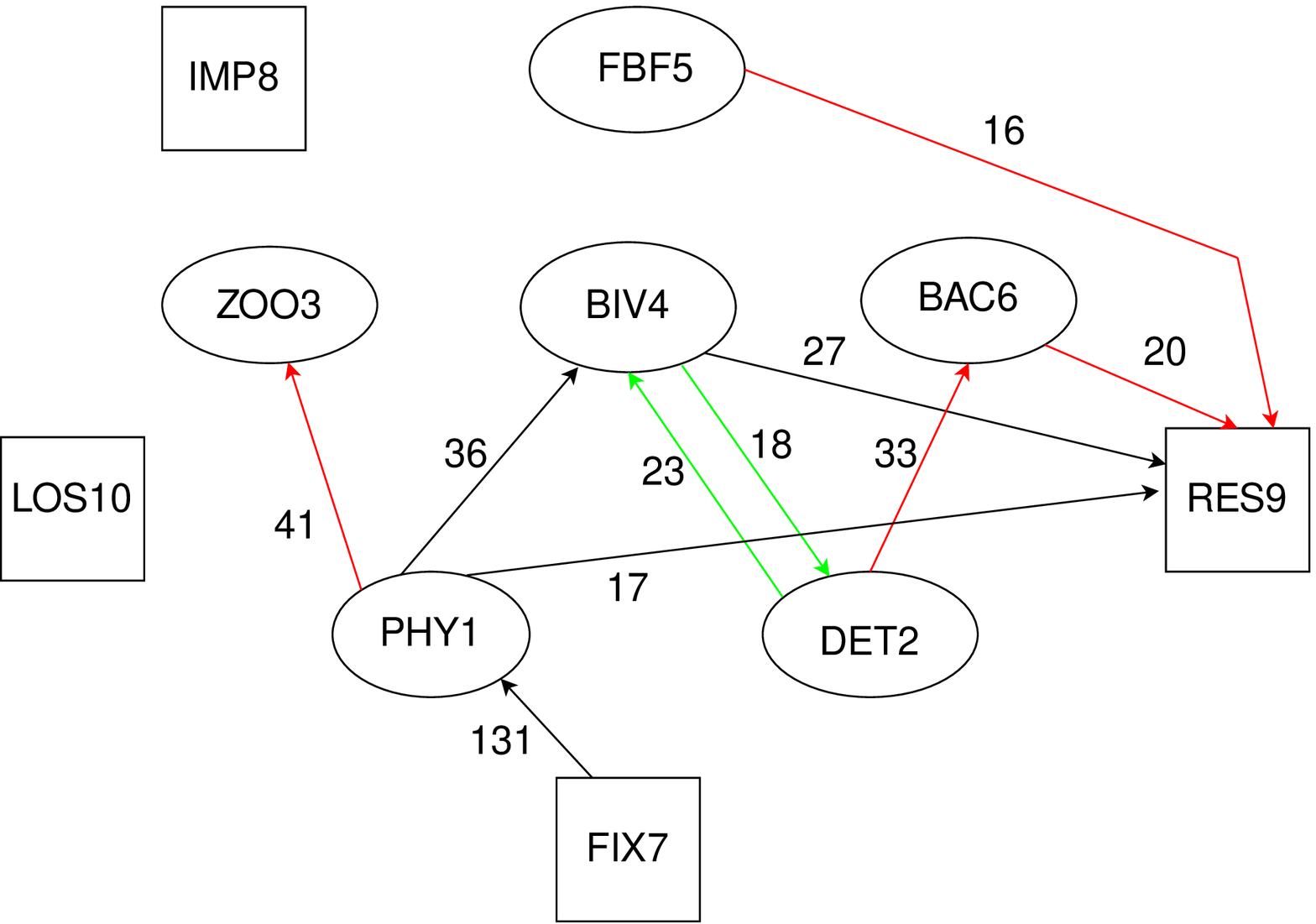,height=6cm,width=5cm}}

\fbox{\epsfig{file = 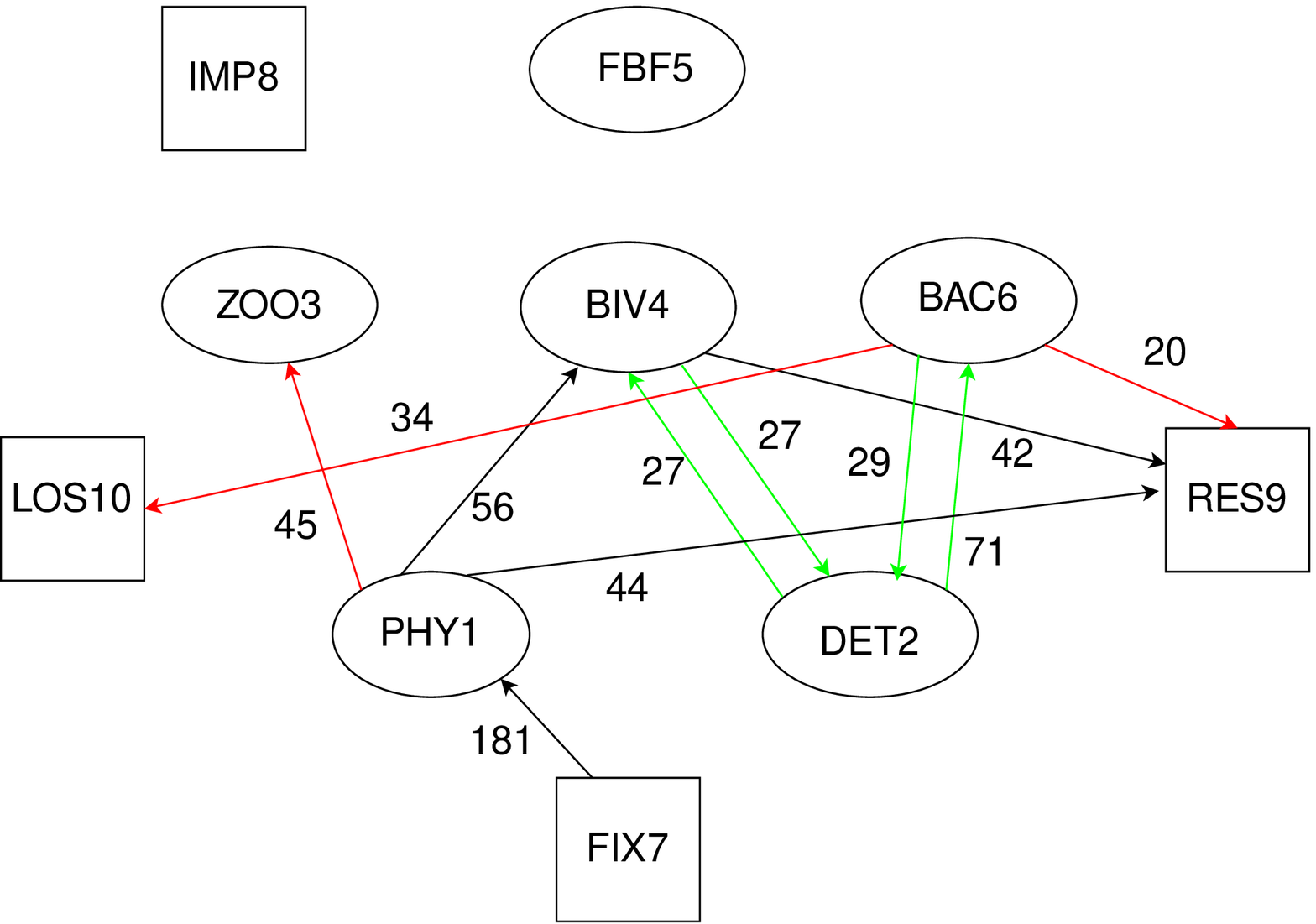,height=6cm,width=5cm}}
\fbox{\epsfig{file = 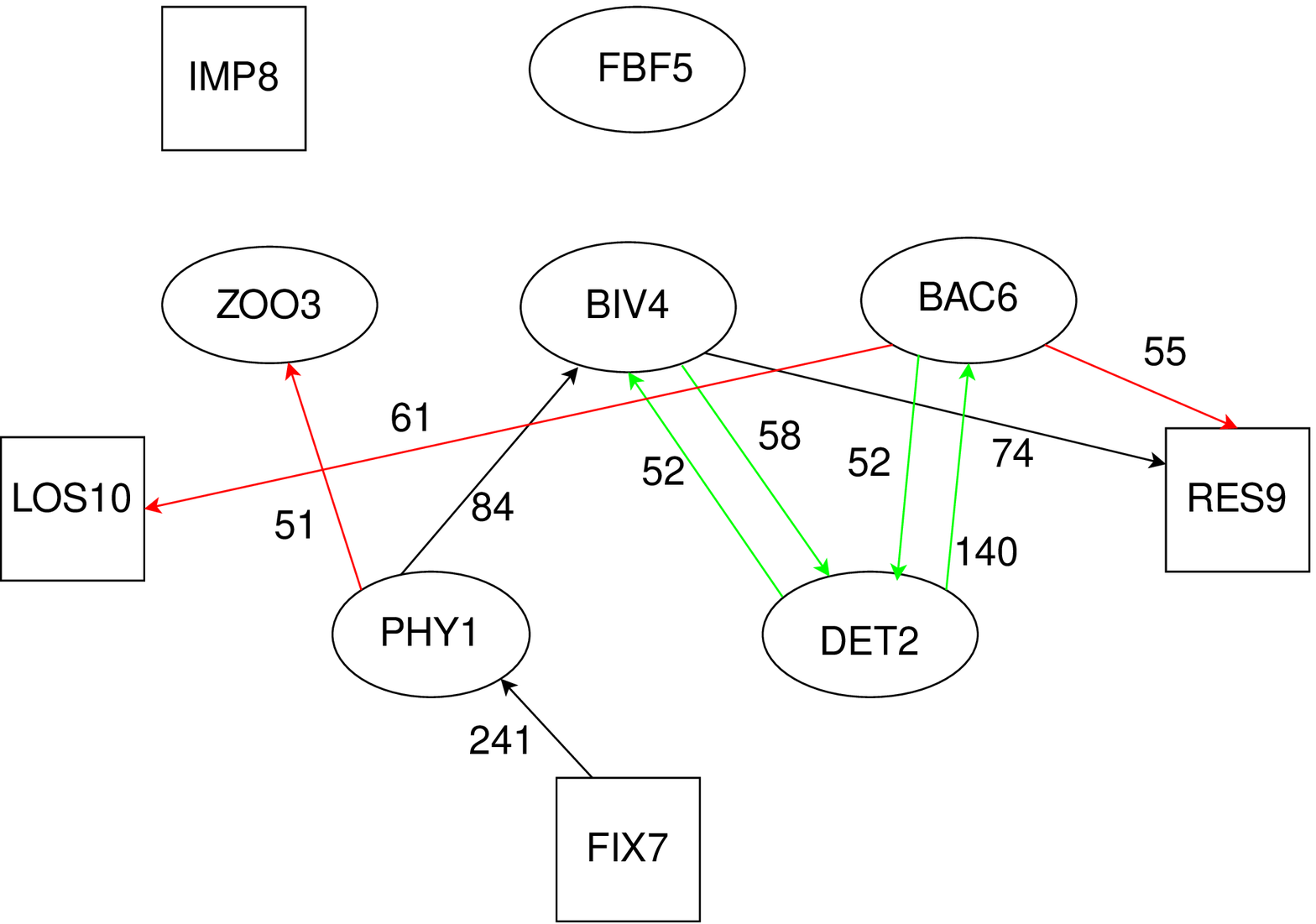,height=6cm,width=5cm}}
\caption{The largest flows for the 6-species system and the goal functions: \textbf{top} $F_1$ (left) and $F_2$ (right), \textbf{middle} $F_3$, \textbf{bottom} $F_4 $ (left) and $F_5$ (right).  Red lines are paths and green lines are  circuits.}
\label{g6f45}
\end{center}
\end{figure}

The flows corresponding to the optimum for the five
different goal functions and the 6-species ecosystem are presented in Table \ref{tab2}.
Using the flow decomposition estimation introduced in the
previous section, we computed the main paths and the circuits
for each flow solution. These are shown {respectively in black and green}
in Fig. \ref{g6f45}. 

For instance for the graph   minimizing $F_{1}$,  the main   paths/circuits are:
\begin{eqnarray}
& \mbox{FIX7} \xrightarrow{33} \mbox{PHY1} \xrightarrow{33} \mbox{LOS10},&\nonumber \\
& \mbox{FIX7} \xrightarrow{29} \mbox{PHY1} \xrightarrow{29} \mbox{BIV4}\quad \xrightarrow{29} \mbox{RES9},&\label{flufix} \\
&\mbox{BIV4} \xrightarrow{17} \mbox{DET2} \xrightarrow{17} \mbox{BIV4}.& \nonumber
\end{eqnarray}

 Fig.  \ref{g6f45} shows the graph with the largest obtained flows. 
The flows obtained for $F_4$ and $F_5$ appear as the most expected 
results as the largest flow components are located at the 
low trophic levels, corresponding to the classical pyramidal 
view of energy flows proposed by \cite{lindeman42} and 
 \cite{Hutchinson79}.

Note that  the   common path \eqref{flufix} appears for the five optimization functions,
showing the well-known importance of phytoplankton in coastal 
ecosystems. This path is the main input of energy in the system. 
Pathways with largest flows also display 
the role of the bivalves in the system: they are important 
consumers of phytoplankton and producers of detritus. Their role as  recyclers  shows in solutions for all goal functions. In such
ecosystems, bivalves are not the only recyclers, bacteria also recycle through  the bacterial loop. This  shows in solutions yielded by $F_4$ and $F_5$.
One odd path is the importation path to fish FBF in the solution of $F_2$; its importance   displays a state of the food web where 
the connection to other systems is crucial for the high trophic 
level FBF compartment. This solution corresponds to an
extreme case, where the system is unable to sustain the FBF compartment and  thus relies more on importation.

\section{Consistency of the solution: stability of the ecosystem}

Solving the optimization problem (\ref{fullopt}) for
each goal function $F_i$ yields a series of flows $f$ between 
the species (including detritus) vertices. From these flows, using a set of biological rules to be given below, we obtain 
a formal dynamical system where the variables are the biomasses $B_i$.
This system involves coupling coefficients $\alpha$  that will differ 
for the different goal functions.

All these systems have, by construction, the same fixed point 
corresponding to the given biomasses. 
Then, by examining the   eigenvalue with maximal real part of  the Jacobians of these different dynamical systems at the fixed point, we can evaluate the structural stability of the ecosystem for this set of biomasses.

\subsection{The dynamical system}

Each flow solution $f$ of the optimization problem (\ref{fullopt})
can be used to derive a system of differential equations satisfied
by the biomasses.
The following rules will be used to derive these systems of differential
equations.

 If living organisms of species $i$ eat organisms of (either living or detritus) species $j$, 
the flow $f_{i,j}$ is given by a law of mass action or Lotka-Volterra 
coupling -- see \cite{murray},
 \be\label{rule1}
f_{i,j} =  \alpha_{i,j} B_i ~B_{j}.
 \ee

  Other flows are assumed to be proportional to the biomass of the 
start vertex, with a coefficient of proportionality depending also 
on the finish vertex,
\be\label{rule2}
f_{i,j} =  \alpha_{i,j} B_i.  
\ee
 
Thus, the dynamics of the system is given by the  flow $f=(f_{i,j})$, that is  to say by  $\alpha=(\alpha_{i,j})$.   

For the 6-species ecosystem of Fig. \ref{eco6}, the  formal rules \eqref{rule1} and \eqref{rule2}  lead to the   system of differential equations 
\begin{eqnarray}\label{SD1}
     \dot{B_{1}} &=& f_{7,1} - f_{1,9} - f_{1,2} - f_{1,4} - f_{1,3} - f_{1,6} - f_{1,10}, \\
     \dot{B_{2}} &=& f_{1,2} + f_{6,2} + f_{3,2} + f_{4,2} + f_{5,2} - f_{2,4} - f_{2,3} - f_{2,6}, \\
     \dot{B_{3}} &=& f_{2,3} + f_{1,3} - f_{3,9} - f_{3,2} - f_{3,5} - f_{3,4} - f_{3,10},\\
     \dot{B_{4}} &=& f_{2,4} + f_{1,4} + f_{3,4} - f_{4,9} - f_{4,2} - f_{4,5} - f_{4,10}, \\
     \dot{B_{5}} &=& f_{8,5} + f_{3,5} + f_{4,5} - f_{5,9} - f_{5,2} - f_{5,10}, \\
     \dot{B_{6}} &=& f_{2,6} + f_{1,6} - f_{6,9} - f_{6,2} - f_{6,10}.\label{SD6}
\end{eqnarray}

By construction, the fixed point of the dynamical system is obtained at the known biomasses of the 6-species ecosystem, 
\begin{equation}\label{Bconnu}
B_{1}^0 = 3.24 ,\ B_{2}^0 = 19 ,\  B_{3}^0 = 1.72 ,\  B_{4}^0 = 19.5,\  B_{5}^0 = 3.19 ,\  B_{6}^0 = 0.75.
\end{equation}

At the equilibrium point $\dot{B_{i}} = 0$ for all $ i$, and we 
get, using  \eqref{SD1}-\eqref{SD6},
\begin{eqnarray*}\label{systab1}
     \alpha_{7,1}B_{1} - \alpha_{1,9}B_{1} - \alpha_{1,2}B_{1} - \alpha_{1,4}B_{1}B_{4} - \alpha_{1,3}B_{1}B_{3} - \alpha_{1,6}B_{1}B_{6} - \alpha_{1,10}B_{1} &=& 0, \\\label{systab2}
     \alpha_{1,2}B_{1} +\hphantom{ml\alpha_{6,2}B_{6} + \alpha_{3,2}B_{3} + \alpha_{4,2}B_{4} + \alpha_{5,2}B_{5} - \alpha_{2,4}B_{2}B_{4} - \alpha_{2,3}B_{2}B_{3}} &&\\
     \alpha_{6,2}B_{6} + \alpha_{3,2}B_{3} + \alpha_{4,2}B_{4} + \alpha_{5,2}B_{5} - \alpha_{2,4}B_{2}B_{4} - \alpha_{2,3}B_{2}B_{3} - \alpha_{2,6}B_{2}B_{6}& =& 0 ,\\
     \alpha_{2,3}B_{2}B_{3} + \hphantom{\alpha_{6,2}B_{6} + \alpha_{3,2}B_{3} + \alpha_{4,2}B_{4} + \alpha_{5,2}B_{5} - \alpha_{2,4}B_{2}B_{4} - \alpha_{2,3}B_{2}B_{3}} &&\\
      \alpha_{1,3}B_{1}B_{3} - \alpha_{3,9}B_{3} - \alpha_{3,2}B_{3} - \alpha_{3,5}B_{3}B_{5} - \alpha_{3,4}B_{3}B_{4} - \alpha_{3,10}B_{3} &=& 0, \\
     \alpha_{2,4}B_{2}B_{4} + \hphantom{\alpha_{6,2}B_{6} + \alpha_{3,2}B_{3} + \alpha_{4,2}B_{4} + \alpha_{5,2}B_{5} - \alpha_{2,4}B_{2}B_{4} - \alpha_{2,3}B_{2}B_{3}} &&\\
      \alpha_{1,4}B_{1}B_{4} + \alpha_{3,4}B_{3}B_{4} - \alpha_{4,9}B_{4} - \alpha_{4,2}B_{4} - \alpha_{4,5}B_{4}B_{5} - \alpha_{4,10}B_{4} &=& 0, \\
     \alpha_{8,5}B_{5} + \alpha_{3,5}B_{3}B_{5} + \alpha_{4,5}B_{4}B_{5} - \alpha_{5,9}B_{5} - \alpha_{5,2}B_{5} - \alpha_{5,10}B_{5} &=& 0, \\
     \alpha_{2,6}B_{2}B_{6} + \alpha_{1,6}B_{1}B_{6} - \alpha_{6,9}B_{6} - \alpha_{6,2}B_{6} - \alpha_{6,10}B_{6} &=& 0. \label{systab6}
\end{eqnarray*}

\subsection{Stability of the fixed point }\label{5.2 sec}

A standard way to compute the stability of the fixed point is
to evaluate the eigenvalues of the Jacobian matrix of the
system. If the real part of one or more eigenvalues is positive,
then the fixed point is unstable.

For the 6-species system,  the Jacobian  is
\begin{equation}
 J=\hphantom{mmmmmmmmmmmmmmmmmmmmmmmmmmmmmmmmm} \label{jacobian}
 \end{equation}
{\small
 $$
\begin{bmatrix} 0 & 0 & -\alpha_{1,3}B_{1} & -\alpha_{1,4}B_{1} & 0 & -\alpha_{1,6}B_{1} \\ \alpha_{1,2} & -\alpha_{2,4}B_{4}-\alpha_{2,3}B_{3}-\alpha_{2,6}B_{6} & \alpha_{3,2}-\alpha_{2,3}B_{2} & \alpha_{4,2}-\alpha_{2,4}B_{2} & \alpha_{5,2} & \alpha_{6,2}-\alpha_{2,6}B_{2} \\ \alpha_{1,3}B_{3} & \alpha_{2,3}B_{3} & 0 & -\alpha_{3,4}B_{3} & -\alpha_{3,5}B_{3} & 0 \\ \alpha_{1,4}B_{4} & \alpha_{2,4}B_{4} & \alpha_{3,4}B_{4} & 0 & -\alpha_{4,5}B_{4} & 0 \\ 0 & 0 & \alpha_{3,5}B_{5} & \alpha_{4,5}B_{5} & 0 & 0 \\ \alpha_{1,6}B_{6} & \alpha_{2,6}B_{6} & 0 & 0 & 0 & 0 \end{bmatrix}
$$
}

Using the optimal solutions obtained in Table \ref{tab2} for the different goal functions, the biomass values \eqref{Bconnu}, and rules \eqref{rule1} and \eqref{rule2}, 
we can compute the maximal real part of the eigenvalues of $J$, shown in Table \ref{tab3}.  
\begin{table} 
\centering
\begin{tabular}{|l|c|c|c|c|c|}
 \hline
 Cost function& $F_{1}$ & $F_{2}$ &$F_{3}$ & $F_{4}$ &$F_{5}$\\ 
 \hline
  $\max(\mathcal{R}e(\lambda))$ & 0.093   & 0.021  & 0.104  & 0.149   & -0.193 \\
 \hline
\end{tabular}
\caption{Maximum values of the real parts of the eigenvalues  of the Jacobian associated to the five goal functions for the  6-species ecosystem.}
\label{tab3}
\end{table}
 
\begin{figure}
\begin{center}
\epsfig{file = 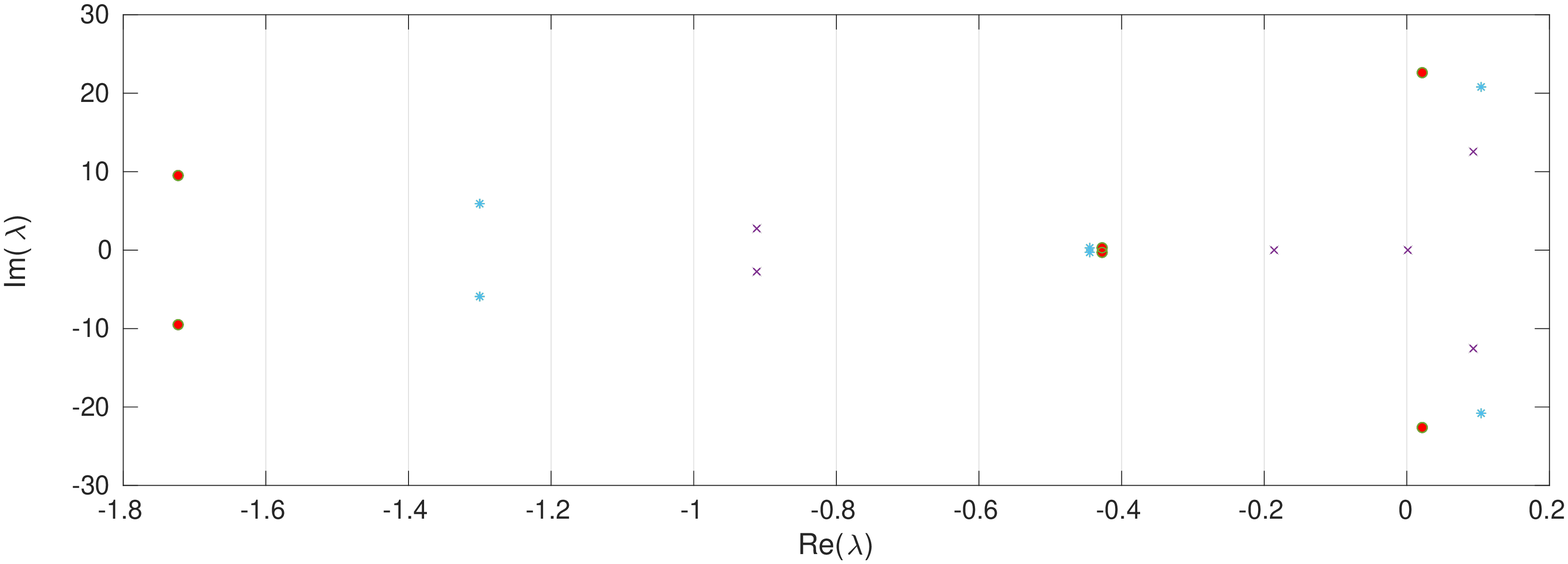,height=4cm,width=10 cm}

\epsfig{file = 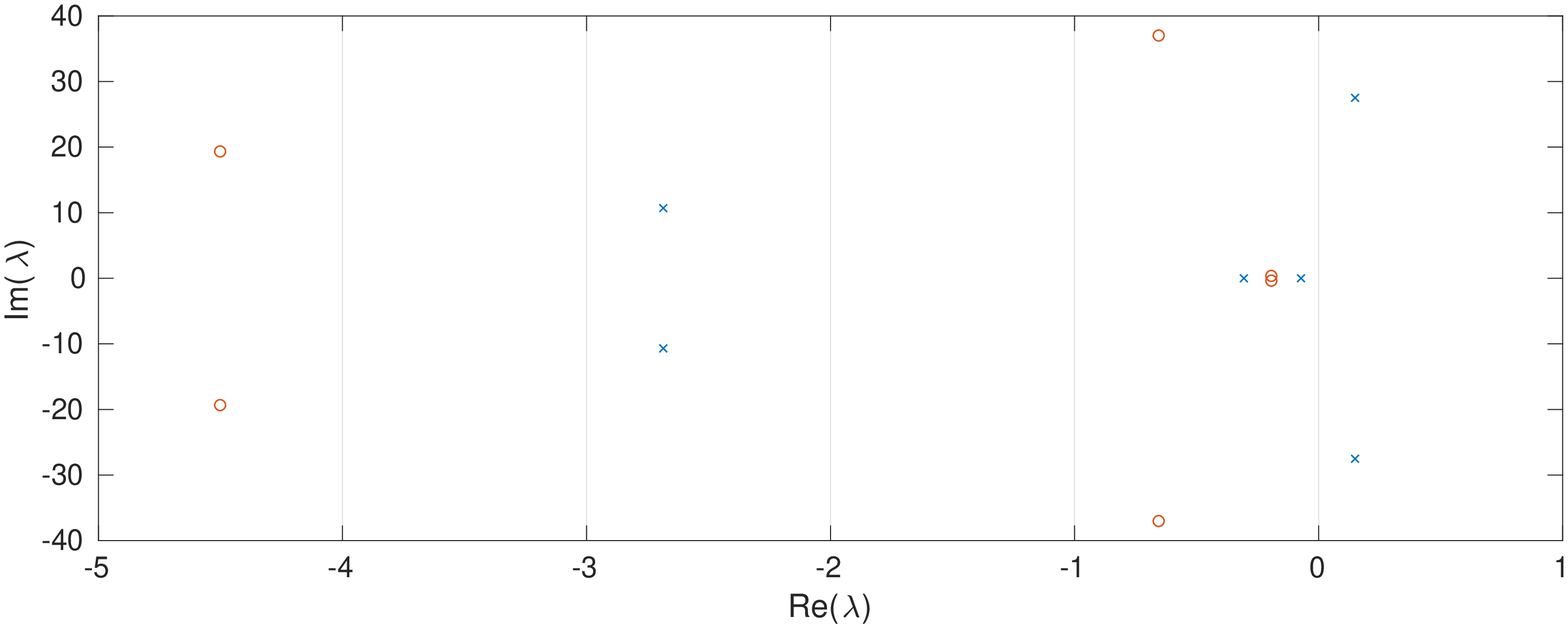,height=4cm,width=10 cm} 
\caption{The spectra in the complex plane of the Jacobians for
the goal functions:  \textbf{top} $F_1$ (black $\times$), 
$F_2$ (red o),    $F_3$  (blue *); \textbf{bottom} $F_4$ (black $\times$), $F_5$ (red o). }
\end{center}
\label{j45}
\end{figure}
Table \ref{tab3} shows that only $F_{5}$ yields a stable biomass fixed point.
Further, Fig.~4 shows the spectra in the complex plane of the 
Jacobians for the five goal functions. Again, the systems are 
unstable for $F_1$ to $F_4$. For all four cases, the eigenvalues 
with maximum real part are close and have a large imaginary part 
indicating strong oscillations.
These goal functions give rise to similar dynamical behaviors.
On the contrary, $F_5$ gives rise to a weakly stable
fixed point with no oscillations.   
There are then at least two main regions in the parameter space, giving 
different behaviors.

\section{Ecosystem dynamics with or without detritus }

From line 2 of the Jacobian $J$ in \eqref{jacobian}, we see that the detritus acts in a different way than the regular species. To understand this effect, we will detail the equations governing the dynamical systems with and without detritus.

\subsection{Lotka-Volterra models for the ecosystem without and with detritus}
First, we assume that the detritus species is absent so that we have
a pure Lotka-Volterra coupling, of which the theory is well established. Since the considered ecosystem involve no coupling $B_i-B_i$ , a Lyapunov function
can be found to show that the equilibrium point is (globally)
stable. We will follow 
\cite{lu79}, see also 
\cite{goh77}.

Consider the Lotka-Volterra model for $n$ species without detritus 
$$
\frac{dB_{i}}{dt} = B_{i}\left (\beta_{i}+\sum_{j=1}^{n} \alpha_{ji}B_{j}\right ) , \quad i = 1,\dots,n,
$$
where $\alpha_{ij} = -\alpha_{ji} \quad \mbox{for all }  i,j$
for the couplings (\ref{rule1})
and $\beta_{i}= \sum_j \alpha_{ji}$ for the couplings (\ref{rule2}).

The nontrivial equilibrium 
$(\tilde{B}_{1},\dots,\tilde{B}_{n})$ is the solution 
of the system 
$$\beta_{i}+\sum_{j=1}^{n} a_{ji}B_{j} = 0, \quad i = 1,\dots,n.
$$
This equilibrium  is feasible, that is $ \tilde{B}_{i} > 0 ,$ for all $i .$
Moreover, $A$ is a skew-symmetric matrix so that $A+A^{T} = 0$, where $A^T$ denotes the transpose of $A$.  
These two conditions induce that 
 the Lotka-Volterra model is {{globally} } stable, 
see \cite{lu79}.

Thus, the detritus is what will determine the stability.
Its in and out flows depend on the goal function.

%
%
%
%
%
%
%
%

Let us now consider the system with detritus.
 The Lotka-Volterra model with $n$ species including detritus is
\begin{eqnarray*}
 \frac{dB_{i}}{dt} &=& B_{i}\Big(\beta_{i}+\sum_{j=1}^{n} \alpha_{ji}B_{j}\Big) , \quad i\neq 2, \\
\frac{dB_{2}}{dt} &=& \sum_{i=1, i\neq 2}^{n} B_{i}(\alpha_{i2} -  \alpha_{2i}B_{2}),
\end{eqnarray*}
where $B_{2}$ is the biomass of the  detritus, $\alpha_{ij} = -\alpha_{ji}$ for all $i,j\neq 2$, $\alpha_{i2}\ge 0$ and $ \alpha_{2i}\ge 0$   for all $  i\neq 2$.

The Jacobian matrix of this system is
\be\label{jj}
     J=J^{0}+R,
 \ee
where 
$ J^{0}_{ij} = \alpha_{ji}B_{i} $ for all $i,j$,  $ R_{2i} = \alpha_{i2} $,  and $ R_{22} = -\sum_{i} \alpha_{2i}B_{i} $, 	and $ R_{ij} = 0 $ for  $  i\neq2 $.
One can see that 
\be\label{dd}
J^{0} = DA,
\ee
where $D$ is the diagonal  matrix of biomasses, with $d_{jj}=B_j$, and $A$ is a skew-symmetric matrix.

For example, for the six species system, we have

$$
J_0 ={\small \begin{bmatrix} 0 & 0 & -\alpha_{1,3}B_{1} & -\alpha_{1,4}B_{1} & 0 & -\alpha_{1,6}B_{1} \\ 
 0 & 0  & -\alpha_{2,3}B_{2} & -\alpha_{2,4}B_{2} & 0  & -\alpha_{2,6}B_{2} \\ \alpha_{1,3}B_{3} & \alpha_{2,3}B_{3} & 0 & -\alpha_{3,4}B_{3} & -\alpha_{3,5}B_{3} & 0 \\ \alpha_{1,4}B_{4} & \alpha_{2,4}B_{4} & \alpha_{3,4}B_{4} & 0 & -\alpha_{4,5}B_{4} & 0 \\ 0 & 0 & \alpha_{3,5}B_{5} & \alpha_{4,5}B_{5} & 0 & 0 \\ \alpha_{1,6}B_{6} & \alpha_{2,6}B_{6} & 0 & 0 & 0 & 0 \end{bmatrix} }
 = D  A,$$ 
where 
$$D ={\small  \begin{bmatrix} 
B_1 & 0 & 0 & 0 & 0 & 0 \\
0   & B_2 & 0 & 0 & 0 & 0 \\
0   &0   &  B_3 & 0 & 0 & 0 \\
0   & 0   &0   &  B_4 & 0 & 0 \\
0   &0   &0   &0   &    B_5 & 0 \\
0   &0   &0   &0   &0   &   B_6 
\end{bmatrix}}
\quad\mbox{and}\quad
A ={\small  \begin{bmatrix}
0 & 0 & -\alpha_{1,3} & -\alpha_{1,4} & 0 & -\alpha_{1,6} \\ 
0 & 0  & -\alpha_{2,3} & -\alpha_{2,4} & 0  & -\alpha_{2,6} \\ 
\alpha_{1,3} & \alpha_{2,3} & 0 & -\alpha_{3,4} & -\alpha_{3,5} & 0 \\ 
\alpha_{1,4} & \alpha_{2,4} & \alpha_{3,4} & 0 & -\alpha_{4,5} & 0 \\ 
0 & 0 & \alpha_{3,5} & \alpha_{4,5} & 0 & 0 \\ 
\alpha_{1,6} & \alpha_{2,6} & 0 & 0 & 0 & 0 \end{bmatrix} } .
$$
Finally, the matrix $R$ is 
$$R ={\small \begin{bmatrix}
0 &  0 & 0  &   0 &  0  &  0 \\
\alpha_{12} &  -\alpha_{24} B_4 -\alpha_{23} B_3 -\alpha_{26} B_6  & \alpha_{32} & \alpha_{42} & \alpha_{52} & \alpha_{62} \\
0 &  0 & 0  &   0 &  0  &  0 \\
0 &  0 & 0  &   0 &  0  &  0 \\
0 &  0 & 0  &   0 &  0  &  0 \\
0 &  0 & 0  &   0 &  0  &  0 
\end{bmatrix}}.
$$
 
 \subsection{Sufficient condition for stability}\label{CSSTAB}
 
Evaluating how  the detritus   will change the stability of the system is a difficult problem.
Nevertheless,   perturbation theory can help to understand how 
eigenvalues of $J_{0}$ get displaced to the ones of $J = J_{0} + R$ when the norm of $R$ is much smaller than the norm of $J_0$, say  $||R|| \ll ||J_0||$; see \cite{scott}. The 
standard complex inner product on $\mathbb{C}^{n}$ will be denoted by $(.,.)$ and  
${\overline z}$ is the conjugate of $z$.

\begin{proposition}With the notation of the previous section,
a sufficient condition for the stability of the system with detritus is
\be \label{csstab}
     \mathcal{R}e(v^{i},Rv^{i}) \leq 0, \quad   i\in S,
\ee
where the $v^{i}$ are eigenvectors  associated to the eigenvalues   $\lambda_0^i$ of  $J^{0}$.
\end{proposition}
{\bf proof} We will prove that \eqref{csstab} implies that the real parts of the  eigenvalues of $J$ are all negative.

An approximation $ \lambda_{J}^i$ of an eigenvalue of $J$ is given by
\begin{eqnarray}
    & \lambda_{J}^i = \lambda_0^{i}+\frac{(w^i ,R v^{i})}{(w ,v^{i})}, \label{l1}
\end{eqnarray}
see \cite{scott}, where  the vector $w $   satisfies 
\be 
   J_{0}^T w = \overline{\lambda_0^i} w  . \label{omega}
\ee

In our special context, the eigenvalues of $J_{0}= D A$ are zero or pure
imaginary (see Lemma \ref{spectrum-infinite}  in  Appendix 3), so that without detritus, the fixed point of the
biomasses is marginally stable. The presence of the detritus will shift 
this stability.
Let us   prove that $w  = D^{-1}v^{i}$ satisfies 
 \eqref{omega}.
 
Indeed, since D is invertible, we deduce from 
$J_{0} v^i= DAv^{i} = \lambda_0^iv^{i} $
that 
$ Av^{i} = \lambda_0^iD^{-1}v^{i} .$

Since $J_{0}^{T} = (DA)^{T} = -AD$, we compute
\be  J_{0}^Tw  = -ADD^{-1}v^{i} = -Av^{i} = -\lambda_0^iD^{-1}v^{i}=-\lambda_0^iw.
\ee
 Thanks to \eqref{dd},  
 Lemma \ref{spectrum-infinite}  applies to show that  $ -\lambda_0^i={\overline \lambda_0^i} $.
Therefore, $J_{0}^Tw   =\overline{\lambda_0^i}w , $ and $w  = D^{-1}v^{i}$ is an eigenvector of $J_{0}^T$, satisfying \eqref{omega}.

Thus, according to \eqref{l1},
$$ \lambda_{J}^i = \lambda_0^i + \frac{(D^{-1}v^{i},Rv^{i})}{(D^{-1}v^{i},v^{i})}.$$
The real parts of $\lambda_0^i$ are all null, so  $\mathcal {R} e(\lambda_{J}^{i}) \leq 0 $ as soon as
\begin{eqnarray*}
    & \Delta_{i} \equiv 
\mathcal {R} e\left(\frac{(D^{-1}v^{i},Rv^{i})}{(D^{-1}v^{i},v^{i})}\right ) \leq 0.
\end{eqnarray*}

Since 
$$(D^{-1}v^{i},v^{i}) = \sum_{j=1}^{n} \frac{|v^i_{j}|^{2}}{B_{j}} > 0\quad \mbox{for all }  i, $$
we only need to consider the sign of the real part of
$$ (D^{-1}v^{i},Rv^{i}) = \frac{1}{B_{2}}(v^{i},Rv^{i}).
$$
Then
$ \Delta_{i} \leq 0 $ if and only if $ \mathcal{R}e(v^{i},Rv^{i}) \leq 0$, and  \eqref{csstab} is indeed a sufficient condition for the stability of the ecosystem.

\section{Discussion and conclusion}

Our methodology can easily be applied to general realistic ecosystems
with a larger number of species and flow components. To show
this, let us present the analysis of the 19-species system defined and studied in \cite{nogues20}.

\begin{figure}
\begin{center}
\fbox{\epsfig{file = 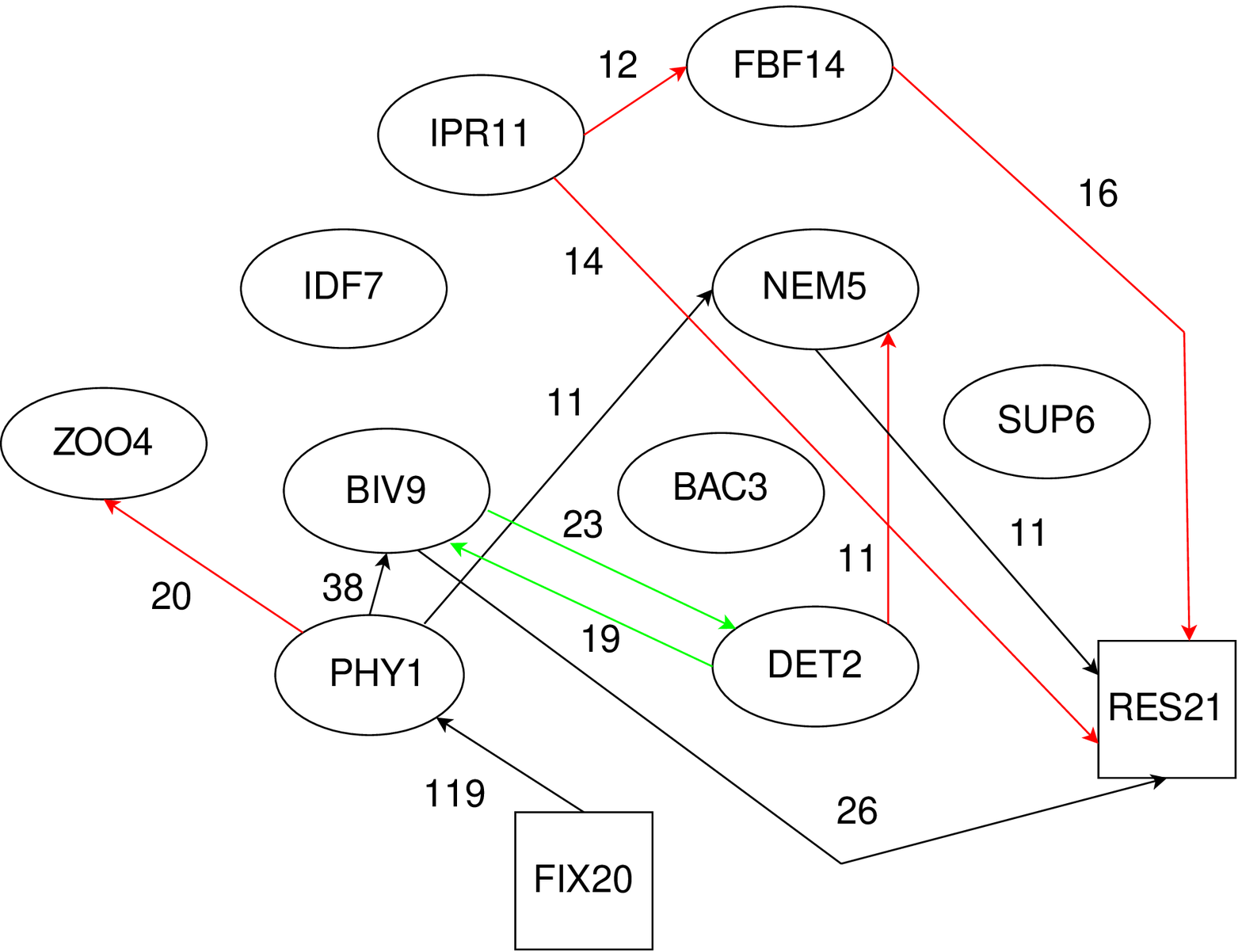,height=6cm,width=5cm}}
\fbox{\epsfig{file = 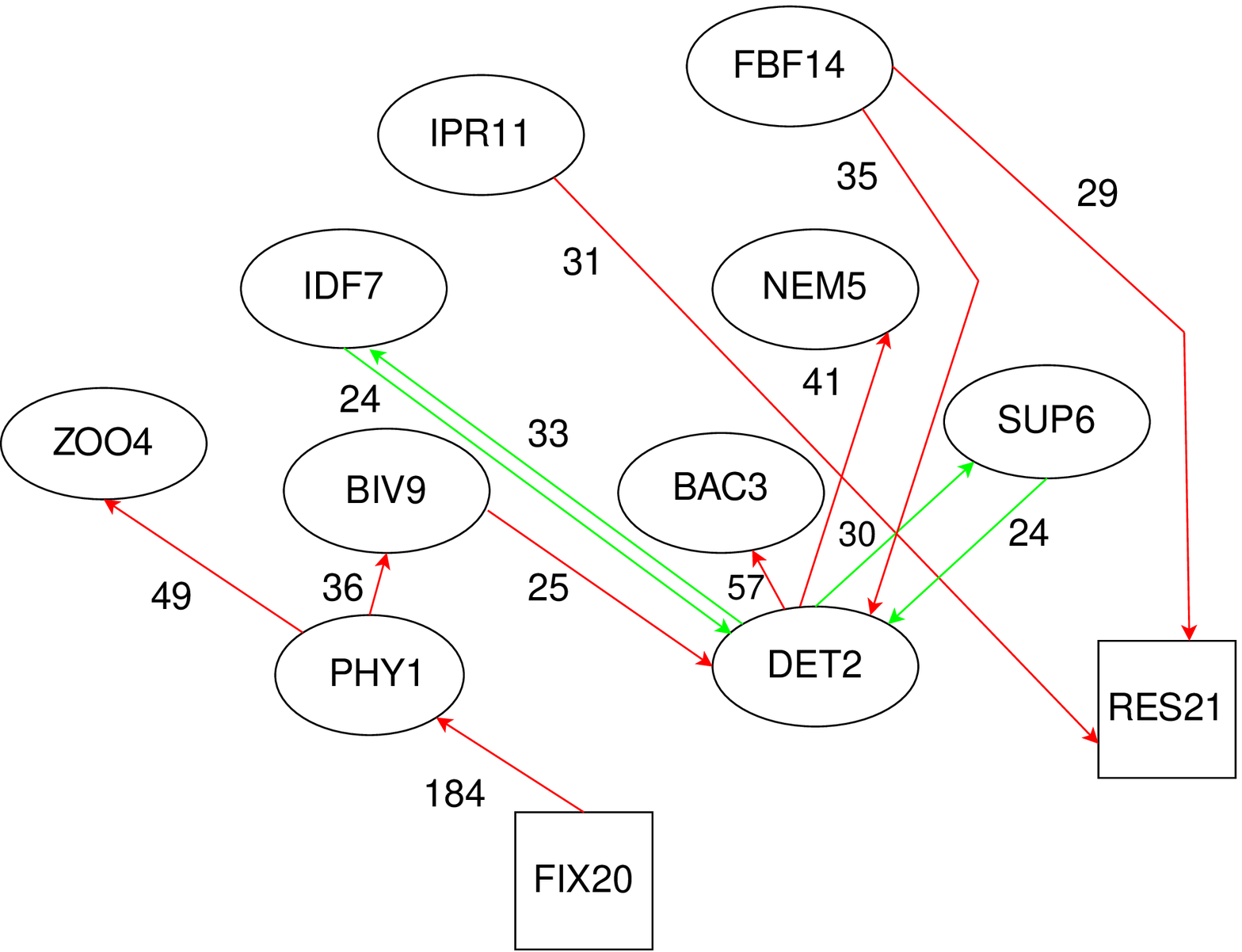,height=6cm,width=5cm}}
 \\
\fbox{\epsfig{file = 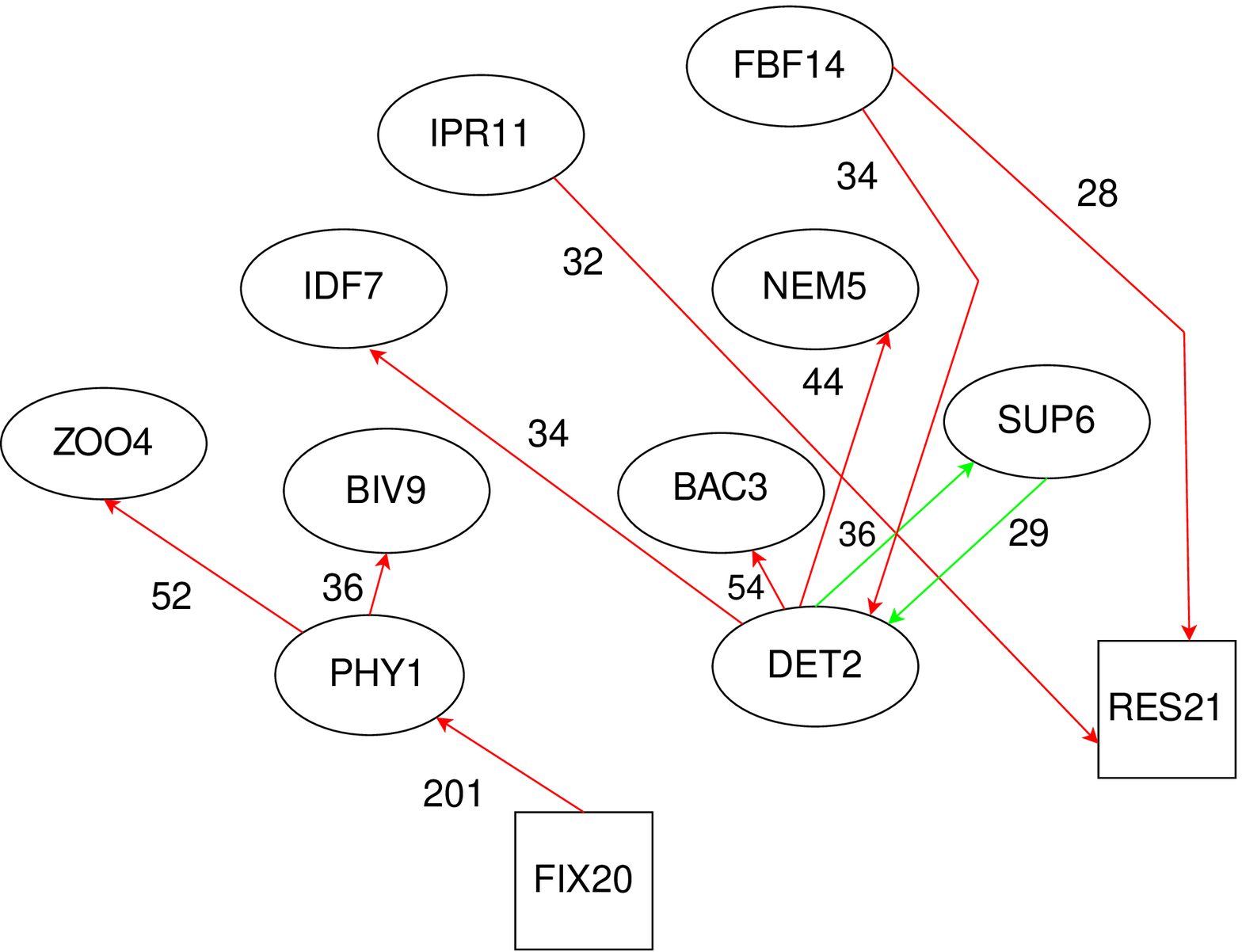,height=6cm,width=5cm}}
 \\
\fbox{\epsfig{file = 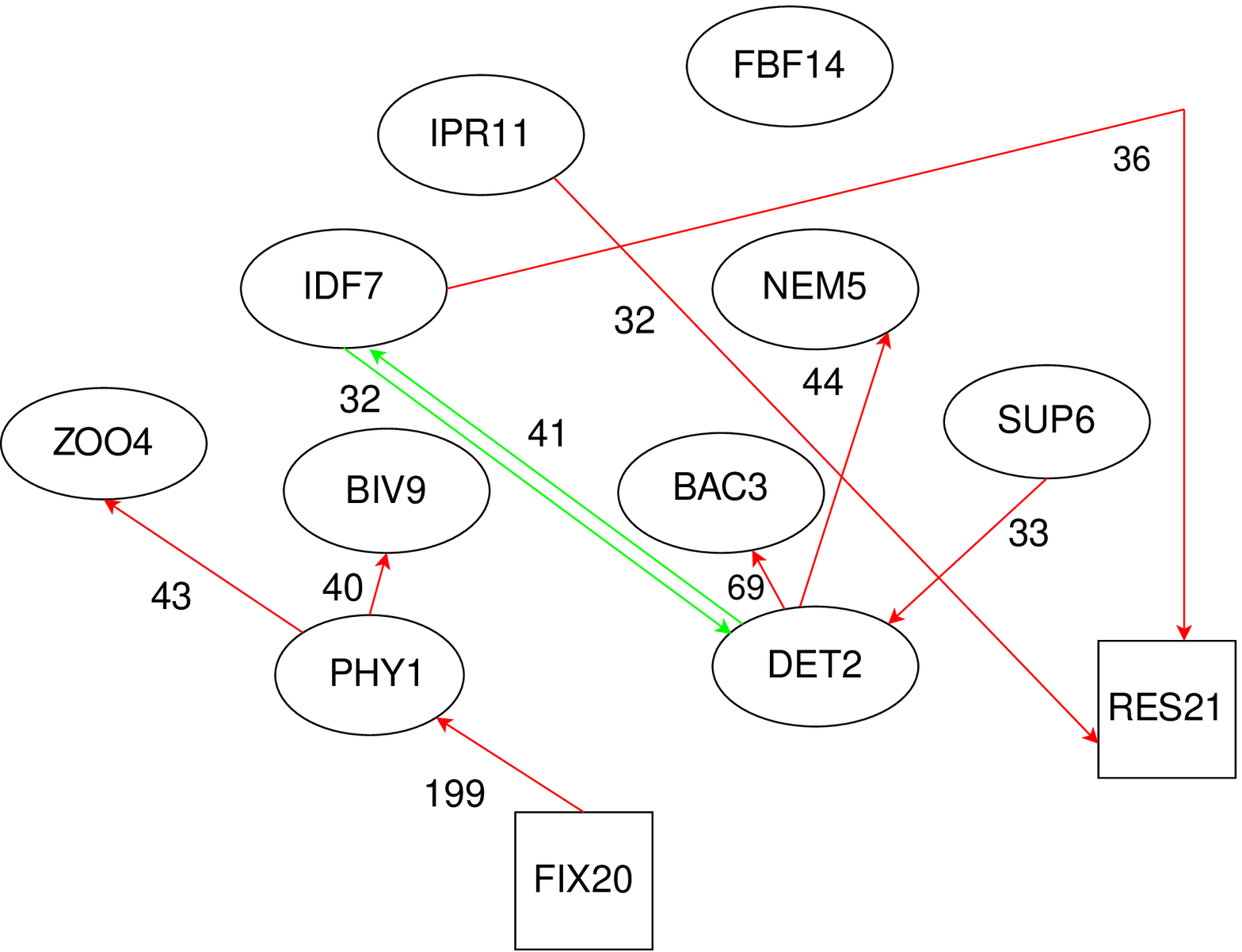,height=6cm,width=5cm}}
\fbox{\epsfig{file = 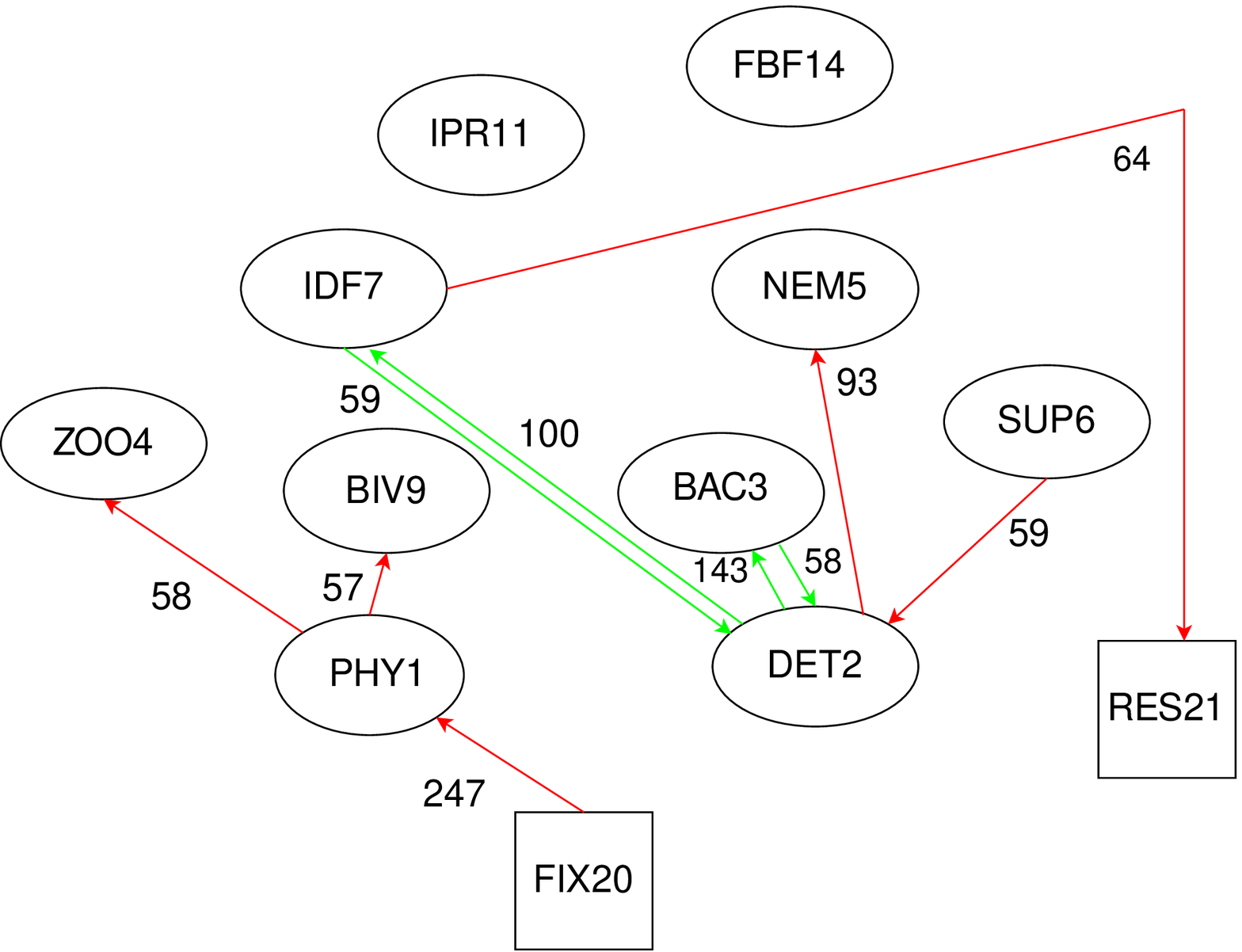,height=6cm,width=5cm}}
\caption{The largest flows for the goal functions:  \textbf{top} $F_1$ (left) and $F_2$ (right), \textbf{middle} $F_3$,  \textbf{bottom}
$F_4$ (left) and $F_5$ (right).}
\label{g19f45}
\end{center}
\end{figure}

Fig. \ref{g19f45} shows the largest flow components
similarly to Fig. \ref{g6f45}.
As for the 6-species system, ZOO3 and BIV4 are the principal 
phytoplankton-eating animals in the 19 species system.

We built for each solution of the
optmisation problem and different goal function a 
dynamical system and estimate the stability of the fixed point.
The maximum of the real parts of the eigenvalues of the Jacobians
are reported in Table \ref{tab4}, similar to Table 3. All the flows correspond to an unstable fixed point showing that the constraints
may need to be refined.

\begin{table}  
\centering
\begin{tabular}{|l|c|c|c|c|c|}
 \hline
 Goal function &$F_{1}$ & 
 $F_{2}$&  $F_{3}$& $F_{4}$ &$F_{5}$ 
\\\hline 
$\max(\mathcal{R}e(\lambda))$ &
 0.075   & 0.127  & 0.242 & 0.126 & 0.233 \\
 \hline
\end{tabular}
\caption{Maximum values of the real parts of the eigenvalues  of the Jacobian associated to the five goal functions for the 19-species ecosystem.}
\label{tab4}
\end{table}

We also examined the ecosystems studied in \cite{ula18}.
For all the cases, the biomass fixed point was found to be stable. 
For many ecosystems, as for example the Crystal River Creek in \cite{ula18}, 
the detritus component has a large biomass. This induces stability, as 
was shown above.

Above methods may also be applied to human environmental networks, or in economics, for example to economic resource trade flow networks; see \cite{kharazi13}, \cite{huang}, and the references therein.

To conclude, we introduced a methodology to study the possible 
flows of an ecosystem defined by observational biomass data and realistic
biological constraints. We formalized the constraints and described
precisely the polytope containing the solutions.
We presented a convex optimization problem based on ecological network
indices used as goal functions. The method is fast, can be used
for large systems and provides a solution within the polytope according
to the indices.

The minimal flow for each goal function can be analyzed using two
different complementary tools. First, the flow is decomposed
into principal paths and circuits. These enable ecologists
to discriminate between the different solutions. Second,
   the consistency of the flow is examined by introducing a
dynamical system and studying the stability of the biomasses fixed point.

\section{Appendix}
\subsection*{Appendix 1: SQL local algorithm of Section \ref{4.2sec}}

The SQL local algorithm solves a nonlinear minimization problem under constraints, say:
\begin{eqnarray*}
    & \min_f \quad F(f),  \\
    & h_{i}(f) = 0 ,\ i=1,\dots, m ,\\
    & g_{j}(f) \leq 0,\ j =1,\dots, n.
\end{eqnarray*}

 Let $\mathcal{L}(f,\alpha,\beta)=F(f)-\alpha^{T} h(f) + \beta^{T} g(f)
$ denote  the  Lagrangian function of this problem where
$\alpha$  and $\beta$ are Lagrange multipliers of dimensions $m$
and $n$ respectively.
Let $\mathcal{W}$ denote  the Hessian matrix of $\mathcal{L}$, defined by 
\begin{eqnarray*}
    & \mathcal{W}_{k} \equiv \mathcal{W}(f_{k},\alpha_{k},\beta_{k})=\bigtriangledown_{ff}^{2} \mathcal{L}(f_{k},\alpha_{k},\beta_{k}),
\end{eqnarray*}
 and $\mathcal{A}(f)$ the Jacobian matrix of the constraints, 
\begin{eqnarray*}
    & \mathcal{A}(f)^{T}=[\bigtriangledown h_{1}(f),\dots, \bigtriangledown h_{m}(f),\bigtriangledown g_{1}(f),\dots,\bigtriangledown g_{n}(f)].
\end{eqnarray*}

Since the cost function $F$ is convex, $\mathcal{W}_{k}$ is a positive definite 
matrix and  $\mathcal{A}(f)$ is a full rank matrix.
Then, at an iterate $f_{k}$, a basic sequential quadratic programming algorithm defines an appropriate search direction 
$p_{k}$ as a solution to the quadratic programming subproblem
\begin{eqnarray*}
    & \min_p \quad \frac{1}{2}p^{T}\mathcal{W}_{k}p+\bigtriangledown f_{k}^{T}p, \\
    & \bigtriangledown h_{i}(f_{k})^{T}p+h_{i}(f_{k})=0,\quad i=1,\dots,m, \\
    & \bigtriangledown g_{j}(f_{k})^{T}p+g_{j}(f_{k})=0,\quad j =1,\dots,n.
\end{eqnarray*}
This is solved by the following algorithm:
\begin{enumerate}
\item Take the initial points $(f_{0},\alpha_{0},\beta_{0})$;

\item {\bfseries For} k = 0,\ 1,\ 2,\ 3, \dots

\quad  Evaluate $F_{f_k},$ $ \bigtriangledown F_{f_k},$ $ \mathcal{W}_{k}=\mathcal{W}(f_{k},$ $\alpha_{k},$ $\beta_{k}),$ $ h_{k},$ $g_{k},$ $ \bigtriangledown h_{k}$ and $\bigtriangledown g_{k};$

\quad  Solve the quadratic subproblem for obtaining $p_{k},\alpha_{k}, \beta_{k};$

\quad  $f_{k+1} \longleftarrow f_{k}+p_{k};$ $ \alpha_{k+1} \longrightarrow u_{k};$ $ \beta_{k+1} \longrightarrow v_{k};$

\quad  {\bfseries If} the condition of convergence is satisfied;

\quad \quad  {\bfseries STOP} with the approximate solution;
      
\item {\bfseries End(For).}
\end{enumerate}
The computing time on an Intel I7 processor and number 
of function calls are given in Table
5.  The code will be made available on github.

\begin{table} 
\centering
\begin{tabular}{|l|c|c|}
 \hline
 Goal function & Computing time & function evaluations \\
 \hline
 $F_{1}$ & 0.2s & 127 \\
 $F_{2}$ & 0.3s & 2002 \\
 $F_{3}$ & 0.4s & 2118 \\
 $F_{4}$ & 0.3s & 2901 \\
 $F_{5}$ & 0.2s & 307 \\
 \hline
\end{tabular}
\label{tab5}
\begin{tabular}{|l|c|c|}
 \hline
 Goal function & Computing time & function evaluations \\ 
 \hline
 $F_{1}$ & 0.9s & 583 \\
 $F_{2}$ & 7s & 55684 \\
 $F_{3}$ & 12s & 60905 \\
 $F_{4}$ & 9s & 78447 \\
 $F_{5}$ & 1s & 1742 \\
 \hline
\end{tabular}
\caption{Computing times and function evaluations: \textbf{top} 6-species system; \textbf{bottom}  19-species system}
\end{table}

\subsection*{Appendix 2: Convexity of goal functions }

In this appendix, we show that all the  goal functions considered in Section \ref{goal} are
convex.

 Both $F_{1}$ and $F_{5}$ are   sums of squares so they are obviously convex. 
Further, if a function is twice continuously differentiable, then it is 
convex if and only if its Hessian is positive semidefinite. We therefore calculate the Hessians $\mathcal{H}_{i}$ of 
$F_{i} $  for $i=$2, 3, 4. First
$$
\mathcal{H}_{2} =\mathcal{H}_{4} = \begin{bmatrix} \frac{1}{p_{i,j}} & 0 & ... & 0 \\ 0 & \frac{1}{p_{k,l}} & ... &  0 \\ ... & ... & ... & ... \\ 0 & 0 & ... & \frac{1}{p_{r,s}} \end{bmatrix} ,
$$
that is obviously positive definite. This shows that indeed the entropy$F_{2}$ and the Kullback Leibler divergence $F_{4}$ are convex.

Let us rewrite the redundancy  $F_{3}$ as
$$
     F_{3} = \sum_{(i,j)\in E} p_{i,j}\ln\left (\frac{p_{i,j}}{p_{i.}}\right ) + \sum_{(i,j)\in E} p_{i,j}\ln\left (\frac{p_{i,j}}{p_{.j}}\right ).
$$
The goal is to show that each term $T^1_{ij} \equiv p_{i,j}\ln(\frac{p_{i,j}}{p_{i.}})$ is convex.
The Hessian   of $T_{ij}$ is
\begin{gather*}
\mathcal{H}_{ij} = \begin{bmatrix} \frac{(p_{i.}-p_{i,j})^{2}}{p_{i,j}p_{i.}^{2}} & -\frac{p_{i.}-p_{i,j}}{p_{i.}^{2}} & ... & -\frac{p_{i.}-p_{i,j}}{p_{i.}^{2}} \\ -\frac{p_{i.}-p_{i,j}}{p_{i.}^{2}} & \frac{p_{i,j}}{p_{i.}^{2}} & ... &  \frac{p_{i,j}}{p_{i.}^{2}} \\ ... & ... & ... & ... \\ -\frac{p_{i.}-p_{i,j}}{p_{i.}^{2}} & \frac{p_{i,j}}{p_{i.}^{2}} & ... & \frac{p_{i,j}}{p_{i.}^{2}} \end{bmatrix} 
\end{gather*}
Let us use the  Sylvester's criterion to show that $T_{ij}$ is 
convex; see \cite{Sylvester}. This criterion says  that a Hermitian matrix is  positive semidefinite if and only if all of its leading principal minors are positive. 

The Hessian $\mathcal{H}_{ij}$ is a Hermitian matrix. Let $M_{ij}^k$ denote the principal minors of this Hessian. We have :
$$ M_{ij}^{1} =  \frac{(p_{i.}-p_{i,j})^{2}}{p_{i,j}p_{i.}^{2}} \geq 0    \quad\mbox{and}\quad 
     M_{ij}^{2} = \mbox{det} \begin{bmatrix} \frac{(p_{i.}-p_{i,j})^{2}}{p_{i,j}p_{i.}^{2}} & -\frac{p_{i.}-p_{i,j}}{p_{i.}^{2}}  \\ -\frac{p_{i.}-p_{i,j}}{p_{i.}^{2}} & \frac{p_{i,j}}{p_{i.}^{2}}  \end{bmatrix} = 0 
.$$
Moreover, $M_{ij}^k = 0$ for all $k \geq 3$, because then $M_{ij}^k$ has at least two equal lines (or columns). 
Therefore, the   Sylvester's criterion is satisfied and 
each term $T^1_{ij}$ is convex.

Similarly, each term $T^2_{ij} \equiv p_{i,j}\ln(\frac{p_{i,j}}{p_{.j}})$ 
is also convex, and hence $F_{3}$ is convex as a sum of convex terms.

Let us now show, using a counter-example
that the ascendency, defined by \cite{hu84} as
$$
 A(f)= \sum_{ij\in E}  \frac{f_{i,j}}{f_{..}}\ln \left(\frac{f_{i,j}f_{..}}{f_{i.}f_{.j}}\right) 
 $$
 is not convex.  
In this aim, consider the graph with three flows $f_{12}, ~f_{23}$ and $ f_{13}$ shown on Fig. 6.
\begin{figure} 
\begin{center}
  \includegraphics[scale = 0.6]{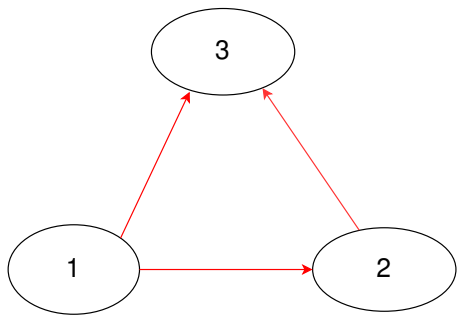}
\caption{Example of 3-species model ecosystem}
\end{center}
\label{eco3}
\end{figure}

We  compute
$$A=f_{13}\,\ln \left({{f_{13}}\over{\left(f_{13}+f_{12}\right)\,
 \left(f_{23}+f_{13}\right)}}\right)-f_{23}\,\ln \left(f_{23}+f_{13}
 \right)-f_{12}\,\ln \left(f_{13}+f_{12}\right).
$$
The gradient of $A$ is given by
\begin{align*} 
& A_{f_{12}}=-\ln \left(f_{13}+f_{12}\right)-1, \qquad
A_{f_{23}}= -\ln \left(f_{13}+f_{23}\right)-1, \\
& A_{f_{13}}=\ln \left({{f_{13}}\over{f_{13}^2+\left(f_{23}+f_{12}\right)\,f_{13}+f_{12}\,f_{23}}}\right)-1.
\end{align*}
There is only one extremum, 
$f_{12}=f_{23}=0 $ and $f_{13}=e^{-1} \approx 0.3678 .$

The Hessian of $A$ is
$${\mathcal H}_A = \begin{pmatrix} -{{1}\over{f_{13}+f_{12}}}&0&-{{1}\over{f_{13}+f_{12}}}\cr 0&-{{1}\over{f_{13}+f_{23}}}&
 -{{1}\over{f_{13}+f_{23}}}\cr -{{1}\over{f_{13}+f_{12}}}&-{{1}\over{f_{13}+f_{23}}}&-{{f_{13}^2-f_{12}\,f_{23}
 }\over{f_{13}^3+\left(f_{23}+f_{12}\right)\,f_{13}^2+f_{12}\,f_{23}\,f_{13}}}\cr \end{pmatrix}. $$
For $f_{12}=1,f_{23}=1,f_{13}=0.4$, we get 
\begin{equation*}
H =\begin{pmatrix}
-0.714 &0&-0.714 \cr 0&-
 0.714 &-0.714 \cr -0.714 &-
 0.714 &1.0714 \cr
\end{pmatrix} ,
\end{equation*}
which is clearly not positive semidefinite.

\subsection*{Appendix 3:  Lemma \ref{spectrum-infinite}}

A classical result is that the eigenvalues of a skew-symmetric real 
matrix are pure imaginary or zero. Indeed, let $A$ be a skew-symmetric 
matrix and $B = iA$, then $B^{*} = -i A^{T} = iA = B$ and 
therefore $B$ is Hermitian. Since $B$ has all real eigenvalues 
${\lambda_{1},..., \lambda_{n}}$, all the eigenvalues of $A$ 
are of the form $-i \lambda_1, \dots, -i \lambda_n$ and thus all pure imaginary.

The following modified version deserves to be proven.  
\begin{lemma}
\label{spectrum-infinite}
Let $J^0$ be a matrix such that $J^0= DA$,  where $D$ is diagonal with non zero elements and
$A$ is skew-symmetric.  Then the  eigenvalues of $J^0$ are either imaginary numbers  or zero.
\end{lemma}

{\bf proof} \\
Let $\lambda$ be an eigenvalue of $J^0$. Let $v$ be an associated   eigenvector,  such that $ D A v = \lambda v$. 
Since $D$ is invertible, we have
\be\label{eq1} A v = \lambda D^{-1} v .
\ee

First, the product of both sides of \eqref{eq1}  with ${\overline v}^T$  gives
\be\label{eq1a}
{\overline v}^T A v = \lambda {\overline v}^T D^{-1} v .
\ee
 Since $\lambda {\overline v}^T D^{-1} v $  is a scalar, taking  transpose of both sides of \eqref{eq1a} yields {$ (A v)^T {\overline v}= -v^T A {\overline v}$}, and hence
\be\label{eq3}\lambda {\overline v}^T D^{-1} v = (A v)^T {\overline v}= -v^T A {\overline v}. \ee

Second,   the complex conjugate  of \eqref{eq1}  is
$ A {\overline v} = {\overline \lambda} D^{-1} {\overline v} . $
The product of both sides with $v$ gives
\be\label{eq4} v^T A {\overline v} = {\overline \lambda} v^T D^{-1} {\overline v}. \ee

Finally, since $v^T D^{-1} {\overline v} = (D^{-1}{\overline v},\overline v) = (D^{-1}v,v) > 0$, identifying the two expressions of $v^T A {\overline v}$ in 
(\ref{eq3}) and (\ref{eq4}) yields
$\lambda = -{\overline \lambda},$
so that $\lambda$ is a pure imaginary number or zero, and the lemma is proven.

\end{document}